\documentclass[12pt]{article}
\usepackage{amsmath}
\usepackage{graphicx}
\usepackage{natbib}
\usepackage{url} 

\usepackage[nolists]{endfloat}

\newcommand{\blind}{0}

\usepackage{subfiles}

\usepackage{multirow}

\usepackage[top=3.2cm, bottom=3.2cm, left=2.4cm, right=2.4cm]{geometry}

\usepackage{fancyhdr}

\usepackage[bottom]{footmisc}

\usepackage{lineno}

\usepackage{hyperref}
\hypersetup{
    colorlinks=true,
    linkcolor=blue,
    citecolor=blue,
    filecolor=magenta,      
    urlcolor=blue,
}

\usepackage{array}
\newcolumntype{P}[1]{>{\centering\arraybackslash}p{#1}} 

\usepackage{siunitx}
\usepackage{dcolumn}
\newcolumntype{L}{D{.}{.}{2,2}}

\usepackage{bookmark}

\usepackage{bm}

\usepackage{setspace}

\usepackage{amsmath}
\usepackage{amsfonts}
\usepackage{amssymb}
\usepackage{amsthm}

\usepackage[T1]{fontenc}
\usepackage{mathtools}

\usepackage{mathptmx}

\usepackage{textcomp}

\usepackage{color, colortbl}
\usepackage[svgnames]{xcolor}

\usepackage{enumitem}

\usepackage{framed}

\usepackage{cancel}

\usepackage{amsmath}
\usepackage[linesnumbered,ruled]{algorithm2e}

\usepackage{graphicx}
\usepackage{caption}
\usepackage{subcaption}
\usepackage{float} 

\usepackage{chngcntr}
\usepackage{aliascnt}
\usepackage{tikz}
\usetikzlibrary{shapes.misc,calc}
\usepackage{tcolorbox}
\tcbuselibrary{skins,theorems,breakable}

\usepackage{epstopdf}

\let\emph\relax 
\DeclareTextFontCommand{\emph}{\bfseries}

\def\newshortcut#1#2{%
\let#1=\undefined
\newcommand{#1}{#2}}

\DeclareSymbolFont{usualmathcal}{OMS}{cmsy}{m}{n}
\DeclareSymbolFontAlphabet{\mathcal}{usualmathcal}
\DeclareSymbolFontAlphabet{\mathbb}{AMSb}

\newshortcut{\C}{\mathbb{C}}
\newshortcut{\F}{\mathbb{F}}
\newshortcut{\N}{\mathbb{N}}
\newshortcut{\Q}{\mathbb{Q}}
\newshortcut{\R}{\mathbb{R}}
\newshortcut{\T}{\mathbb{T}}
\newshortcut{\Z}{\mathbb{Z}}
\newshortcut{\D}{\mathbb{D}}

\newshortcut{\cA}{\mathcal{A}}
\newshortcut{\cB}{\mathcal{B}}
\newshortcut{\cC}{\mathcal{C}}
\newshortcut{\cD}{\mathcal{D}}
\newshortcut{\cE}{\mathcal{E}}
\newshortcut{\cF}{\mathcal{F}}
\newshortcut{\cG}{\mathcal{G}}
\newshortcut{\cH}{\mathcal{H}}
\newshortcut{\cI}{\mathcal{I}}
\newshortcut{\cJ}{\mathcal{J}}
\newshortcut{\cK}{\mathcal{K}}
\newshortcut{\cL}{\mathcal{L}}
\newshortcut{\cM}{\mathcal{M}}
\newshortcut{\cN}{\mathcal{N}}
\newshortcut{\cO}{\mathcal{O}}
\newshortcut{\cP}{\mathcal{P}}
\newshortcut{\cQ}{\mathcal{Q}}
\newshortcut{\cR}{\mathcal{R}}
\newshortcut{\cS}{\mathcal{S}}
\newshortcut{\cT}{\mathcal{T}}
\newshortcut{\cU}{\mathcal{U}}
\newshortcut{\cV}{\mathcal{V}}
\newshortcut{\cW}{\mathcal{W}}
\newshortcut{\cX}{\mathcal{X}}
\newshortcut{\cY}{\mathcal{Y}}
\newshortcut{\cZ}{\mathcal{Z}}

\newshortcut{\fA}{\mathfrak{A}}
\newshortcut{\fB}{\mathfrak{B}}
\newshortcut{\fC}{\mathfrak{C}}
\newshortcut{\fD}{\mathfrak{D}}
\newshortcut{\fE}{\mathfrak{E}}
\newshortcut{\fF}{\mathfrak{F}}
\newshortcut{\fG}{\mathfrak{G}}
\newshortcut{\fH}{\mathfrak{H}}
\newshortcut{\fI}{\mathfrak{I}}
\newshortcut{\fJ}{\mathfrak{J}}
\newshortcut{\fK}{\mathfrak{K}}
\newshortcut{\fL}{\mathfrak{L}}
\newshortcut{\fM}{\mathfrak{M}}
\newshortcut{\fN}{\mathfrak{N}}
\newshortcut{\fO}{\mathfrak{O}}
\newshortcut{\fP}{\mathfrak{P}}
\newshortcut{\fQ}{\mathfrak{Q}}
\newshortcut{\fR}{\mathfrak{R}}
\newshortcut{\fS}{\mathfrak{S}}
\newshortcut{\fT}{\mathfrak{T}}
\newshortcut{\fU}{\mathfrak{U}}
\newshortcut{\fV}{\mathfrak{V}}
\newshortcut{\fW}{\mathfrak{W}}
\newshortcut{\fX}{\mathfrak{X}}
\newshortcut{\fY}{\mathfrak{Y}}
\newshortcut{\fZ}{\mathfrak{Z}}
\newshortcut{\fa}{\mathfrak{a}}
\newshortcut{\fb}{\mathfrak{b}}
\newshortcut{\fc}{\mathfrak{c}}
\newshortcut{\fd}{\mathfrak{d}}
\newshortcut{\fe}{\mathfrak{e}}
\newshortcut{\ff}{\mathfrak{f}}
\newshortcut{\fg}{\mathfrak{g}}
\newshortcut{\fh}{\mathfrak{h}}
\newshortcut{\fj}{\mathfrak{j}}
\newshortcut{\fk}{\mathfrak{k}}
\newshortcut{\fl}{\mathfrak{l}}
\newshortcut{\fm}{\mathfrak{m}}
\newshortcut{\fn}{\mathfrak{n}}
\newshortcut{\fo}{\mathfrak{o}}
\newshortcut{\fp}{\mathfrak{p}}
\newshortcut{\fq}{\mathfrak{q}}
\newshortcut{\fr}{\mathfrak{r}}
\newshortcut{\fs}{\mathfrak{s}}
\newshortcut{\ft}{\mathfrak{t}}
\newshortcut{\fu}{\mathfrak{u}}
\newshortcut{\fv}{\mathfrak{v}}
\newshortcut{\fw}{\mathfrak{w}}
\newshortcut{\fx}{\mathfrak{x}}
\newshortcut{\fy}{\mathfrak{y}}
\newshortcut{\fz}{\mathfrak{z}}

\newshortcut{\vaprhi}{\varphi}

\newshortcut{\arsinh}{\operatorname{arsinh}}
\newshortcut{\arcosh}{\operatorname{arcosh}}
\newshortcut{\artanh}{\operatorname{artanh}}
\newshortcut{\arcsch}{\operatorname{arcsch}}
\newshortcut{\arsech}{\operatorname{arsech}}
\newshortcut{\arcoth}{\operatorname{arcoth}}
\newshortcut{\sign}{\operatorname{sign}}
\newshortcut{\sinc}{\operatorname{sinc}}
\newshortcut{\Log}{\operatorname{Log}}
\newshortcut{\Ei}{\operatorname{Ei}}
\newshortcut{\Si}{\operatorname{Si}}
\newshortcut{\Ci}{\operatorname{Ci}}

\usepackage{amsmath}

\newshortcut{\d}{\partial}
\newshortcut{\del}{\nabla}
\newshortcut{\grad}{\operatorname{grad}}
\newshortcut{\div}{\operatorname{div}}
\newshortcut{\Re}{\operatorname{Re}}
\newshortcut{\Im}{\operatorname{Im}}
\newshortcut{\PV}{\operatorname{PV}}
\newshortcut{\supp}{\operatorname{supp}}
\newshortcut{\spec}{\operatorname{spec}}
\newshortcut{\dist}{\operatorname{dist}}

\newshortcut{\bigplus}{\mathop{\Large+}}

\newshortcut{\bmu}{\bm{\mu}}
\newshortcut{\bnu}{\bm{\nu}}
\newshortcut{\bbeta}{\pmb{\beta}}
\newshortcut{\btheta}{\bm{\theta}}

\newcommand{\transpose}{^\top}

\usepackage[utf8]{inputenc}
\usepackage{listings}
\usepackage{color}
 
\definecolor{codegreen}{rgb}{0,0.6,0}
\definecolor{codegray}{rgb}{0.5,0.5,0.5}
\definecolor{codepurple}{rgb}{0.58,0,0.82}
\definecolor{backcolour}{rgb}{0.95,0.95,0.92}
 
\lstdefinestyle{mystyle}{
    backgroundcolor=\color{backcolour},   
    commentstyle=\color{codegreen},
    keywordstyle=\color{magenta},
    numberstyle=\tiny\color{codegray},
    stringstyle=\color{codepurple},
    basicstyle=\footnotesize,
    breakatwhitespace=false,         
    breaklines=true,                 
    captionpos=b,                    
    keepspaces=true,                 
    numbers=left,                    
    numbersep=5pt,                  
    showspaces=false,                
    showstringspaces=false,
    showtabs=false,                  
    tabsize=2
}
 
\def\sinc{\mathop{\rm sinc}\nolimits}

\def\div{\mathop{\rm div}\nolimits}
\def\tr{\mathop{\rm tr}\nolimits}

\def\dist{\mathop{\rm dist}\nolimits}

\def\supp{\mathop{\rm supp}\nolimits}

\newshortcut{\bzero}{\mathbf{0}}
\newshortcut{\bone}{\mathbf{1}}

\DeclareSymbolFont{symbolsC}{U}{txsyc}{m}{n}
\DeclareMathSymbol{\Perp}{\mathrel}{symbolsC}{121}

\newshortcut{\Exp}{\operatorname{Exp}}
\newshortcut{\Uniform}{\operatorname{Uniform}}
\newshortcut{\Poisson}{\operatorname{Poisson}}
\newshortcut{\Binomial}{\operatorname{Binomial}}

\newshortcut{\id}{\operatorname{id}}
\newshortcut{\tr}{\operatorname{tr}}
\newshortcut{\Tr}{\operatorname{Tr}}
\newshortcut{\rank}{\operatorname{rank}}
\newshortcut{\adj}{\operatorname{adj}}

\newshortcut{\psd}{\succeq}
\newshortcut{\pd}{\succ}

\newshortcut{\smallgap}{\vspace{1em}}

\AtBeginDocument{
\setlength\abovedisplayskip{0.5em}
\setlength\belowdisplayskip{0.5em}
}

\setlist{itemsep=0em,topsep=0em}

\newtheorem*{theorem*}{Theorem}

\newtheorem*{lemma*}{Lemma}

\newtheorem*{corollary*}{Corollary}

\newtheorem*{observation*}{Observation}

\newtheorem*{proposition*}{Proposition}

\newtheorem*{claim*}{Claim}

\theoremstyle{definition}

\newtheorem*{assumption*}{Assumption}

\newtheorem*{definition*}{Definition}

\newtheorem*{example*}{Example}

\newtheorem*{exercise*}{Exercise}

\theoremstyle{remark}

\newtheorem*{remark*}{Remark}

\makeatletter

\@addtoreset{theorem}{section}

\definecolor{mygray}{RGB}{215,215,215}
\definecolor{myblue}{RGB}{17,94,140}

\makeatletter
\tcbset{
	mytheorem/.code args={#1#2#3#4}{%
		\refstepcounter{#2}\label{#4}%
		\pgfkeysalso{title={\setbox\z@=\hbox{#1~\csname the#2\endcsname\ }\hangindent\wd\z@\hangafter=1 \mbox{#1~\csname the#2\endcsname\ }#3}}},%
}

\newcommand{\mtcbmaketheorem}[5]{%
	\newtcolorbox{#1}[3][]{#3,mytheorem={#2}{#4}{##2}{#5:##3},##1}%
}
\makeatother

\tcbset{
	simplestyle/.style={
		breakable,
		freelance,
		fonttitle=\bfseries\sffamily
	},
	thmstyle/.style={
		breakable,
		freelance,
		boxrule=2pt,
		width=\linewidth,
		frame code={%
			\path[fill=myblue,draw=myblue!75!black]
			(frame.north west) -- ([xshift=-8pt]frame.north east) --
			([yshift=-8pt]frame.north east) -- (frame.north east|-interior.north east) --
			(frame.north west|-interior.north west) -- cycle;
		},
		interior titled code={
			\path[fill=mygray!80,draw=mygray]
			(frame.west|-interior.north west) -- (frame.east|-interior.north east) --   
			([yshift=8pt]frame.east|-interior.south east) -- 
			([xshift=-8pt]frame.east|-interior.south east) --
			(frame.west|-interior.south west) -- cycle;
		},
		fonttitle=\bfseries\sffamily
	},
	defstyle/.style={
		breakable,
		freelance,
		boxrule=2pt,
		width=\linewidth,
		frame code={%
			\path[top color=myblue!50,bottom color=myblue!50,
			middle color=myblue!50]
			([xshift=8pt]frame.north west) -- ([xshift=-8pt]frame.north east) --
			([yshift=-8pt]frame.north east) -- 
			(frame.north east|-interior.north east) --
			(frame.north west|-interior.north west) -- 
			([yshift=-8pt]frame.north west) -- cycle;
		},
		interior titled code={
			\path[fill=mygray!80]
			(frame.west|-interior.north west) -| 
			([yshift=8pt]frame.east|-interior.south east) -- 
			([xshift=-8pt]frame.east|-interior.south east) -- 
			([xshift=8pt]frame.west|-interior.south west) -- 
			([yshift=8pt]frame.west|-interior.south west) -- cycle;
			\path[fill=myblue] 
			([xshift=0.5\pgflinewidth,yshift=1.5pt]frame.west|-interior.north west) 
			rectangle 
			([xshift=-0.5\pgflinewidth,yshift=-1.5pt]frame.east|-interior.north east);
		},
		fonttitle=\bfseries\sffamily\normalcolor
	}
}

\mtcbmaketheorem{problembox}{Problem}{simplestyle}{theorem}{ex}
\mtcbmaketheorem{theorembox}{Theorem}{simplestyle}{theorem}{ex}
\mtcbmaketheorem{corollarybox}{Corollary}{simplestyle}{theorem}{ex}

\begin{document}

\def\spacingset#1{\renewcommand{\baselinestretch}%
{#1}\small\normalsize} \spacingset{1}


\if0\blind
{
    \title{Massive Parallelization of Massive Sample-size Survival Analysis}
    \author{Jianxiao Yang$^1$, \\Martijn J.~Schuemie$^{2,3}$, \\Xiang Ji$^{6}$, \\Marc A.~Suchard$^{1, 2, 4, 5}$}
    \date{
        $^1$Department of Computational Medicine, David Geffen School of Medicine at UCLA, Los Angeles, CA, USA\\
        $^2$Department of Biostatistics, Fielding School of Public Health at UCLA, Los Angeles, CA, USA\\
        $^3$Janssen Research and Development, Titusville, NJ, USA\\
        $^4$Department of Human Genetics, David Geffen School of Medicine at UCLA, Los Angeles, CA, USA\\
        $^5$VA Informatics and Computing Infrastructure, US Department of Veterans Affairs, Salt Lake City, UT, USA\\
        $^6$Department of Mathematics, Tulane University, New Orleans, Louisiana, USA
    }
    \maketitle
    
%
%
%
    \newpage
} \fi

\if1\blind
{
  \bigskip
  \bigskip
  \bigskip
  \begin{center}
    {\LARGE\bf Title}
\end{center}
  \medskip
} \fi

\bigskip
\begin{abstract}
Large-scale observational health databases are increasingly popular for conducting comparative effectiveness and safety studies of medical products.
However, increasing number of patients poses computational challenges when fitting survival regression models in such studies.
In this paper, we use graphics processing units (GPUs) to parallelize the computational bottlenecks of massive sample-size survival analyses.
Specifically, we develop and apply time- and memory-efficient single-pass parallel scan algorithms for Cox proportional hazards models and forward-backward parallel scan algorithms for Fine-Gray models for analysis with and without a competing risk using a cyclic coordinate descent optimization approach.
We demonstrate that GPUs accelerate the computation of fitting these complex models in large databases by orders of magnitude as compared to traditional multi-core CPU parallelism.
Our implementation enables efficient large-scale observational studies involving millions of patients and thousands of patient characteristics.
The above implementation is available in the open-source \texttt{R} package \texttt{Cyclops} \citep{suchard2013massive}.
\end{abstract}

\noindent%
\vfill

\newpage
\spacingset{1.5} 

\section{Introduction}

Increasing accessibility of large-scale observational health data provides rich opportunities to study comparative effectiveness and safety of medical products, but also poses unprecedented challenges.
Typical administrative claims and electronic health record (EHR) databases now follow tens to hundreds of millions of individuals \citep{hripcsak2016characterizing} with tens of thousands of possible health conditions, drugs and procedures occurring over decades of patient lives \citep{suchard2013massive}.
The massive scales of these databases offer more power for statistical analyses to learn about the effects of these products on health outcomes but also bring taxing computational burden.

The increasing complexity of common statistical models further exacerbated this big-data problem.
For instance, the Cox proportional hazards model and the Fine-Gray model are widely applied in comparative effectiveness and safety studies.
The computational complexity of likelihood evaluation for the Cox model and the Fine-Gray model naively grows quadratically with sample size.
In addition, some form of regularization \citep{madigan2010bayesian} is often needed to achieve parsimonious model selection when entertaining hundreds to thousands of patient characteristics.
This regularization typically requires computationally intensive cross-validation to select the optimal regularization parameter(s), further straining limited computational resources.

One can distribute computationally intensive work to the cloud or dedicated clusters that house multiple central processing units (CPUs) across separate compute nodes that are linked together loosely through an Ethernet or InfiniBand network \citep{holbrook2020massive}.
However, for problems that require communication between nodes, communication latency may become a severe bottleneck.
Since fitting survival models generally requires iterative algorithms, the communication latency costs often overpower the gains from the parallelized work within each iteration.
To minimized communications between CPUs, we often follow coarse-scale parallelism, where programs are split into small number of large tasks \citep{barney2010introduction}.
However, this may result in load imbalance and only achieve ``embarrassingly parallel'' benefits \citep{suchard2010understanding}.
Furthermore, CPU clusters and the cloud can be costly and arcane for many clinical researchers.

Multi-core CPU parallelization is another choice for distributing intensive computational works, as modern CPU chip usually consists of $2$ to $18$ or more cores that can run independently.
One limitation of this architecture is that all cores share a single ``memory bandwidth'', the amount of data that can be written to or read from memory in a given period of time \citep{holbrook2020massive}.
This approach also suffers from the limited number of cores. Thus, multi-core CPU parallelization is often only useful for modest-scale problems.

In contrast, graphics processing units (GPUs) offer a relatively inexpensive and efficient approach for speeding up fine-scale parallel computation.
In fine-scale parallelism, we decompose programs into a large number of small tasks to facilitate load balancing and achieve much higher level of parallelism than coarse grain approaches \citep{barney2010introduction}.
The GPU is an ideal device for fine-scale parallelism because (1) it consists of hundreds to thousands of compute cores and (2) the shared memory architecture of a GPU's coupled thread block allows for threads to communicate and share data among each other at a very high speed.
Finally, GPUs are often conveniently available on standard laptops and desktop computers and can be externally connected to a personal computer.

Accelerating statistical computing via GPUs is an emerging discipline.
As examples, \cite{zhou2010graphics} attain 100-fold speedups with GPUs in high-dimensional optimization.
\cite{suchard2013massive} demonstrate that GPU parallelization achieves one to two orders of magnitude improvement over CPUs for a Bayesian self-controlled case series model.
\cite{beam2016fast} accelerate Hamiltonian Monte Carlo using GPUs by efficient evaluation of their probability kernel and its gradient.
\cite{terenin2019gpu} demonstrate that Gibbs sampling can run orders of magnitude faster than on a CPU.
\cite{holbrook2020massive} apply GPU computing to a Bayesian multidimensional scaling model and deliver more than 100-fold speedups over serial calculations.
\cite{ko2022high} explore the GPU parallelization for proximal gradient descent on modest-sized $\ell_1$ regularized dense Cox regression using PyTorch.
In this paper, we leverage GPU parallelization to the Cox model and the Fine-Gray model through innovative algorithmic mapping that play to the GPU's strengths for accelerating observational studies utilizing massive healthcare data.
Specifically, we identify the computational bottleneck of the Cox model and the Fine-Gray model and take advantage of the cutting-edge GPU-accelerated library CUB \citep{merrill2016single} that navigate this bottleneck.
We further implement our GPU advances in the easy-to-use \texttt{R} package \texttt{Cyclops} \citep{suchard2013massive}.
Our implementation supports a sparse data format, considering that observational healthcare data are generally sparse; the vast majority of patient characteristics are encoded as the presence or absence of some clinical condition, drug exposure, medical procedure or laboratory measurement above or below a cutoff point within given time-frames.
We finally demonstrate that our GPU implementation accelerates the computation of fitting these complex models by order-of-magnitude compared to a similar CPU implementation on multiple GPUs and CPUs with different technical specifications.

\section{Methods}

\subsection{Cox proportional hazards model}\label{2.1}

We first establish notation under a typical survival analysis setting.
Suppose there are $N$ observed individuals available in a study.
For individual $i = 1, \ldots, N$, let $Y_i = \text{min}(T_i, C_i)$ represent their survival time, where $T_i$ and $C_i$ are the time-to-event time and right-censoring time, respectively.
Let $\delta_i$ be the indicator variable such that $\delta_i = 1$ if we observe the event occurrence of individual $i$ and $\delta_i = 0$ if the individual $i$ is censored.
Let $\mathbf{x}_i$ be a $P$-dimensional vector of time-independent covariates for individual $i$.
The survival data then consist of triplets $\{Y_i, \delta_i, \mathbf{x}_i\}_{i = 1}^n$.

The cumulative distribution function of survival times gives the probability that the event of interest has occurred by time $t$, i.e.~$F(t | \mathbf{x}) = \text{Pr}(T \leq t | \mathbf{x})$.
The survival function gives the probability that the event has not occurred by time $t$, i.e.~$S(t) = \text{Pr}(T > t)$.
Then we define the hazard function of time-to-event time as:
\begin{eqnarray}
h(t) = \lim_{\Delta t\to\infty} \frac{\text{Pr}(t \leq T < t + \Delta t | T \geq t)}{\Delta t} = \frac{f(t)}{S(t)},
\end{eqnarray}
where $f(t) = \frac{d}{dt} F(t)$ is the density function of random variable $T$.

Let $\bbeta = (\beta_1, \beta_2, \ldots, \beta_P)\transpose$ be a $P$-dimensional vector of unknown, underlying model parameters.
Assuming survival times $y_1, y_2, \ldots, y_N$ are independent and identically distributed from density $f(y | \bbeta)$ and $\bbeta$ parameterized the survival function $S(y | \bbeta)$, \cite{cox1972regression} proposes a semi-parametric hazard function as the product of an unspecified baseline hazard function $h_0(y_i | \bbeta)$ and an exponential link function of covariates:
\begin{eqnarray}
h(y_i | \bbeta) = h_0(y_i)\exp{\left(\mathbf{x}_i\transpose\bbeta\right)}.
\end{eqnarray}

Parameter estimation of the Cox proportional hazards model follows from the log-partial likelihood
\begin{eqnarray} \label{cox_l}
l_{\text{\tiny partial}}(\bbeta) = \sum_{i = 1}^N\delta_i\left\{\mathbf{x}_i\transpose\bbeta - \log\left[\sum_{r \in R_1(Y_i)}\exp \left(\mathbf{x}_r\transpose\bbeta\right) \right]\right\},
\end{eqnarray}
where $R_1(Y_i) = \{r : y_r \geq Y_i\}$ denotes the set of subjects who are “at risk” for event at time $Y_i$.
Then one often estimates $\bbeta$ by its maximum log-partial likelihood estimator $\hat{\bbeta}_{\text{\small mple}} = \text{arg max}_{\beta}\{l_{\text{\tiny partial}}(\bbeta)\}$.

This log-partial likelihood has a complicated form due to the repeated calculation of the risk sets, and thus brings a high computation burden.
In practice, we need to keep track of the sum of many terms for each subject that usually requires $\mathcal{O}(N^2)$ number of operations and will explode quickly as $N$ increases \citep{kawaguchi2020scalable}.
This computational burden prevents traditional model fitting as there can be millions of observations available in observational health databases.

\subsection{Fine-Gray model}\label{2.2}

The Fine-Gray model \citep{fine1999proportional} generalizes the Cox proportional hazards model for competing risks time-to-event data that consist of more than one type of event.
Unlike the standard survival analysis setting such as under the Cox model where individuals are only susceptible to one type of event during follow-up, competing risks arise when individuals can experience more than one type of event and the occurrence of one type of event will either prevent the occurrence or change the underlying risk of the others \citep{noordzij2013we}.
For individual $i$, competing risks data inherit the definition of event time $T_i$, possible right censoring time $C_i$, event indicator $\delta_i$, and covariates $\mathbf{x}_i$ from the standard survival data setting, and additionally include an event type variable $\epsilon_i$.
Without loss of generality, we assume there are two types of events, where $\epsilon_i = 1$ indicates that $T_i$ refers to the time of primary event and $\epsilon_i = 2$ indicates the competing risk event.

The cumulative incidence function (CIF) for competing risks data describes the probability of failing from the event of interest before the other possible (competing) event.
Under the above setting when $\epsilon = 1$ indicating the event of interest, the CIF and hazards function are defined as:
\begin{gather}
F_1(t | \mathbf{x}) = \text{Pr}(T \leq t, \epsilon = 1 | \mathbf{x})\text{, and} \\
h_1(t | \mathbf{x}) = \lim_{\Delta t\to\infty} \frac{\text{Pr}\{t \leq T < t + \Delta t, \epsilon = 1 | \{T \geq t\} \cup (\{T < t\} \cap \{\epsilon \neq 1\}),  \mathbf{x}\}}{\Delta t} = -\frac{d}{dt}\log\{1 - F_1(t | \mathbf{x})\}.
\end{gather}

To model the covariate effects on $F_1(t | \mathbf{x})$, \cite{fine1999proportional} propose the proportional subdistribution hazards function:
\begin{eqnarray}
h_1(y_i | \bbeta) = h_{10}(y_i)\exp{(\mathbf{x}_i\transpose\bbeta)},
\end{eqnarray}
where $h_{10}(y_i | \bbeta)$ is an unspecified baseline subdistribution hazard, and $\bbeta$ is a P-dimensional vector of model parameters.

Parameter estimation of the Fine-Gray subdistribution proportional hazards model follows from the log-pseudo likelihood
\begin{eqnarray}\label{fg_l}
l_{\text{\tiny pseudo}}(\bbeta) = \sum_{i = 1}^N I(\delta_i\epsilon_i = 1)\left\{\mathbf{x}_i\transpose\bbeta - \log\left[\sum_{r \in R(Y_i)}\hat{w}_r(Y_i)\exp \left(\mathbf{x}_r\transpose\bbeta\right) \right]\right\},
\end{eqnarray}
where $R(Y_i) = \{r : (y_r \geq Y_i)$ or $(y_r < Y_i$ and $\epsilon_r \neq 1)\}$ denotes the risk set at time $Y_i$, and $\hat{w}_r(t)$ is a time-dependent weight based on an inverse probability of censoring weighting (IPCW) technique \citep{robins1992recovery}.
Assuming two event types exist, the risk set $R(Y_i)$ contains two disjoint set: $R_1(Y_i) = \{r : y_r \geq Y_i\}$ and $R_2(Y_i) = \{r : y_r < Y_i \cap \epsilon_r = 2\}$, where $R_1(Y_i)$ is the regular risk set that includes the observations that have an observed time equalling or after $Y_i$ and $R_2(Y_i)$ includes the observations that have observed the competing event before time $Y_i$.
Here we follow the design of weighted score functions for incomplete data with right censoring in \cite{fine1999proportional} for unbiased estimation from the complete-data partial likelihood.
The time-dependent weights are defined as $\hat{w}_r(t) = I(C_r \geq \text{min}(T_r, t))\hat{G}(t) / \hat{G}(\text{min}(Y_r, t))$, where $\hat{G}(t)$ is the Kaplan-Meier estimate of $G(t)$ and $G(t) = \text{Pr}(C > t)$ is the survival function of censoring variable $C$.
Combined, one can estimate $\bbeta$ by its maximum log-partial likelihood estimator $\hat{\bbeta}_{\text{\small mple}} = \text{arg max}_{\beta}\{l_{\text{\tiny pseudo}}(\bbeta)\}$.

\subsection{Statistical regularization}\label{2.3}

Observational healthcare datasets often include a large number of patient characteristics.
For example, administrative claims usually contain information on all drug prescriptions, medical procedures and diagnosis codes for patients, and EHRs generally further contain  demographics, medical history notes, laboratory results, and other health status indications \citep{madigan2014systematic}.
A statistical regularization approach is typical in such high-dimensional data analysis to avoid overfitting.
We can conveniently add a penalty $\pi(\bbeta)$ for $\bbeta$ to the log-partial likelihood of Cox model or the log-pseudo likelihood of Fine-Gray model and estimate $\bbeta$ through these joint penalized likelihoods to achieve regularization.

For $\ell_1$ regularization that shrinks many components of $\bbeta$ to be zero, we define a separable penalty for each dimension $\beta_j$ in $\bbeta$ through
\begin{eqnarray}
\pi(\bbeta) = \sum_j\pi(\beta_j | \gamma_j) = - \sum_j\gamma_j |\beta_j|,
\end{eqnarray}
where the tuning parameters $\gamma_j$ control the degree of regularization for each dimension.
Similarly, one may employ an $\ell_2$ penalty on the dimensions of $\bbeta$, such that:
\begin{eqnarray}
\pi(\bbeta) = \sum_j\pi(\beta_j | \tau_j) = - \sum_j\frac{\beta_j^2}{2\tau_j}.
\end{eqnarray}
Usually one assumes $\gamma_j = \gamma$ $\forall j$ and $\tau_j = \tau$ $\forall j$, and we choose $\gamma$ or $\tau$ through cross-validation.

Note that statistical inference in the context of regularization remains a challenge.
Various standard errors estimators based on the non-parametric bootstrap have been proposed \citep{casella2010penalized, chatterjee2011bootstrapping}, but an approach that is both computationally efficient and statistically valid still remains out of reach.
Since we are focusing on computational bottleneck in this paper, we decide to follow the standard practice of regularization in large-scale and high-dimensional observational health studies \citep{mueller2016factors, shortreed2017outcome, bramante2021metformin} despite the limitations of regularization.

\subsection{Maximum likelihood estimation using cyclic coordinate descent}\label{2.4}

Following \cite{genkin2007large} and \cite{mittal2014high}, we exploit a cyclic coordinate descent (CCD) algorithm to reduce the high-dimensional penalized likelihood optimization down to a large set of simple one-dimensional optimizations \citep{wu2008coordinate}.
This method cycles through each covariate and updates it using a Newton step while holding all other covariates as constants.
The advantage of CCD is it only requires the calculation of scalar gradients and Hessians and avoids the inversion of large Hessian matrices in high-dimensional regression.

Specifically, for each one-dimensional optimization problem, we pick the $\beta_j^{(new)}$ by maximizing $g(\beta_j) = l(\beta_j) + \pi(\beta_j)$ while holding all other $\beta_j$'s unchanged.
The second-order Taylor approximation of the penalized log likelihood at current $\beta_j$ is:
\begin{eqnarray}
g(z) \approx g(\beta_j) + g'(\beta_j)(z - \beta_j) + \frac{1}{2}g''(\beta_j)(z - \beta_j)^2.
\end{eqnarray}
Then the new estimate $\beta_j^{\text{\tiny (new)}}$ falls out
\begin{eqnarray}
\beta_j^{\text{\tiny (new)}} = \beta_j + \Delta\beta_j = \beta_j - \frac{g'(\beta_j)}{g''(\beta_j)}.
\end{eqnarray}

We employ a trust region approach similar to \cite{genkin2007large} to restrict the the step size so that the quadratic remains a reasonable approximation to the objective and improve convergence.
In particular, we update $\beta_j$ during iteration $k$ by
\begin{gather}
\Delta\beta_j^{(k)} = - \frac{g'(\beta_j^{(k)})}{g''(\beta_j^{(k)})}\text{, and} \\
\beta_j^{(k+1)} = \beta_j^{(k)} + \text{sgn}(\Delta\beta_j^{(k)})\min\{|\Delta\beta_j^{(k)}|, \Delta_j^{(k)}\},
\end{gather}
where we update the trust region half-width as $\Delta_j^{(k+1)} = \max\{2|\Delta\beta_j^{(k)}|, \Delta_j^{(k)} / 2\}$, starting with $\Delta_j^{(0)} = 1$.

Note that both the negated log likelihood of the Cox model and the Fine-Gray model are convex in $\bbeta$, as well as the $\ell_1$ and $\ell_2$ penalty terms.
Although the $\ell_1$-norm is nondifferentiable at origin, we can follow the approach of \citet{wu2008coordinate} to compute the directional derivatives along each forward and backward coordinate direction for our objective function.
In particular, we compute the directional derivatives in both directions by plugging in $\text{sgn}(\beta_j) = +1$ and $\text{sgn}(\beta_j) = -1$ when $\beta_j^{(k)} = 0$, and only update $\beta_j$ in a direction when the directional derivative is negative, otherwise we keep $\beta_j^{(k+1)} = 0$.
Since the objective function is convex, it is impossible for both directional derivatives to be negative, but either direction with a negative directional derivative will result in a successful update.
Although we lack of rigorous proof of convergence when employing a trust region, as the induced step sizes fail to meet the strict convergence conditions for this optimization problem \citep{xu2017globally}, we have not observed this issue in our work.

\subsection{Massive parallelization on GPUs}\label{2.5}

Parallelization through clusters and multi-core CPUs exhibits a number of drawbacks that makes these devices ill-suited for massive survival analysis, as discussed in the Introduction.
As such, this paper exploits massive parallelization on GPUs through new fine-scale algorithm decomposition for speeding up large-scale computations.
Here we begin with an overview on GPU computing and summarize its main strengths and weaknesses. 

The modern GPU contains an array of multithreaded streaming multiprocessors (SMs), where hundreds to thousands of work threads execute simultaneously \citep{nickolls2008scalable}.
Many threads group together as a thread block, in which threads communicate through shared memory and cooperate through barrier synchronization.
Thread blocks are further grouped into kernel grids. 
The programmer specifies the number of threads per block and number of blocks forming the grid.
In our code, we program this ensemble via CUDA, a parallel computing platform that allows general-purpose computing on GPUs (GPGPU) using a familiar C-like syntax. 

Understanding the memory hierarchy of GPUs is important for achieving optimal performance for parallel programs.
Each thread has its own set of processor registers and local memory for thread-private variables, which provide the fastest memory access. 
Each thread block has a limited shared memory pool that is only visible to the threads within this block.
All threads also have access to a large high-bandwidth, but off-chip (global) memory embedded on the GPU card.
Shared memory provide high-speed access, while accessing global memory is hundreds of times slower \citep{micikevicius20093d}.

In our implementation, GPUs handle only the most computationally intensive tasks.
When such a task is scheduled, relevant data are first copied from host CPU memory to the global memory on the GPU device.
Then the GPU kernel is launched, which loads data to on-chip memory for defined operations and writes results to global memory.
Finally, results are copied from the device back to the host.

This parallel programming model has several limitations that we should keep in mind.
First, we should minimize data transfer between the host and device because the transfer is extremely slow.
Second, accessing global memory on the GPU is also relatively slow, so we want to minimize the reads from global memory and the writes to global memory.
When we do read or write from global memory, we want sequential threads to access sequential addresses in memory.  
In this manner, the GPU ``coalesces'' multiple memory requests into a smaller number of 128-byte transactions, and we want to over-subscribe each GPU core with multiple threads, so that the cores remain active through thread-context-switching and the latency in memory access can be hidden.
Third, launching a kernel also has overhead on the order of microseconds, so it is preferred to combine a series of kernels into a larger ``fused'' one.
Finally, contemporary GPUs issue single instructions to a ``warp" of 32 threads simultaneously, such that all threads within a warp must execute the same instruction each clock cycle. 
When threads within a warp follow different data-dependent conditional branches, their execution becomes temporally serialized; this can cause a performance penalty.
To avoid this issue, one attempts to minimize the number of diverging branches within a warp.

\subsection{Tree-based parallel algorithms: reduction and scan}\label{2.6}

Here we review two useful building blocks for massively parallel algorithms: \textit{reduction} and \textit{scan}.
Reductions convert an array of elements into a single result.
For example, if the reduction operator is addition, then the reduction takes an array $[a_0, a_1, \ldots, a_{n-1}]$ and returns a single value $\sum_{i=0}^{n-1} a_i$.
Reductions are useful for implementing log-likelihood calculations, since independent samples contribute additively to the model log-likelihood.
Taking an array $[a_0, a_1, \ldots, a_{n-1}]$, the \textit{scan} operation returns the array $[a_0, a_0 + a_1, \ldots, \sum_{i=0}^{n-1} a_i]$.
If we start from the beginning of the input array as above, the resulting array is called a \textit{prefix sum}.
While the resulting array is called a \textit{suffix sum} if we start from the end and proceed towards the beginning.
Scans are useful in accumulating statistics about individuals in the risk set in survival analysis.
Implementing a sequential version of a reduction or scan both require \textasciitilde$n$ additions on an array of length $n$.

\begin{figure}[ht]
     \centering
     \begin{subfigure}[b]{0.47\textwidth}
         \centering
         \includegraphics[width=\textwidth]{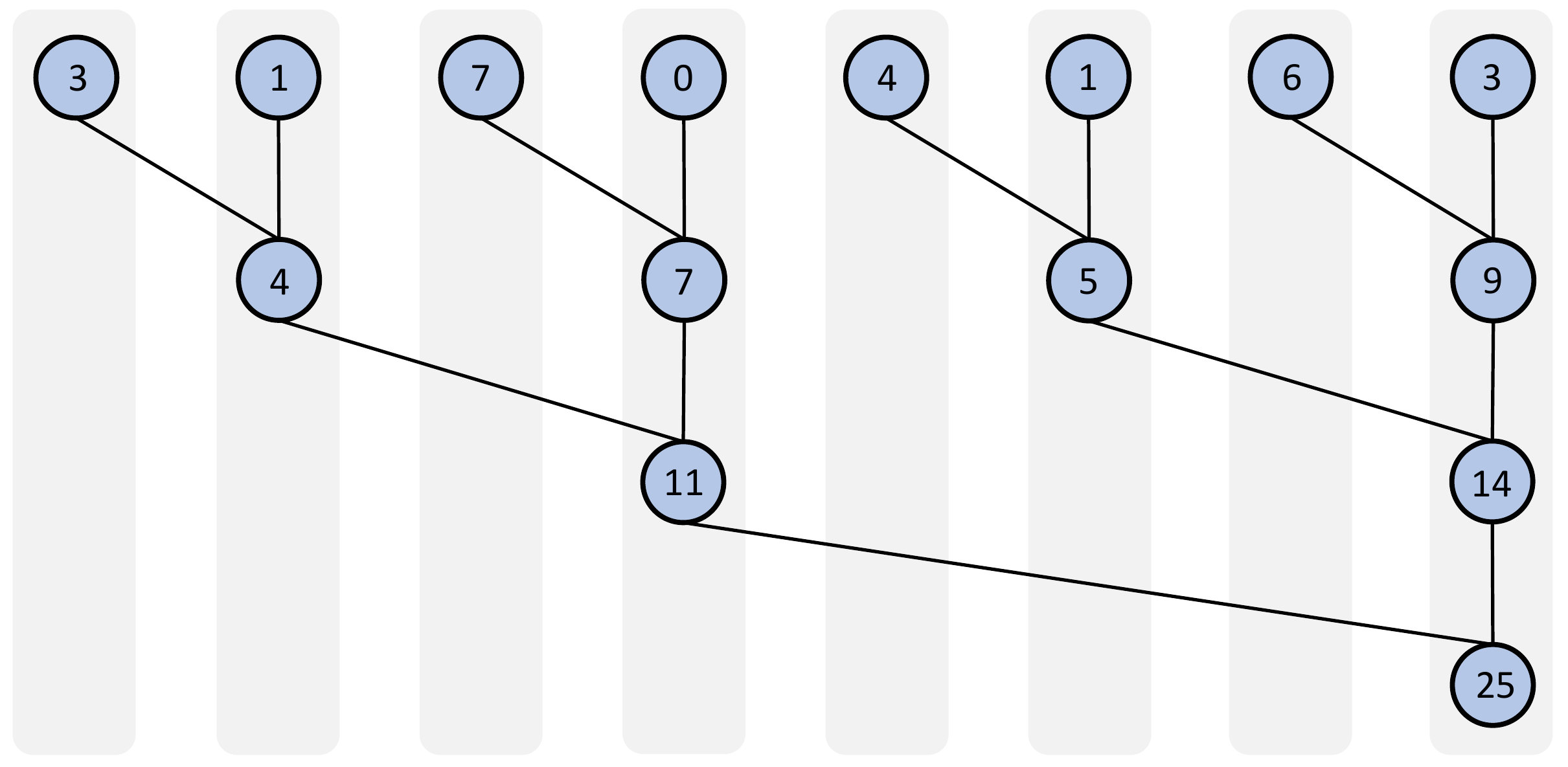}
         \caption{Reduction}
         \label{fig:method_naive_reduce}
     \end{subfigure}
     \hfill
     \begin{subfigure}[b]{0.47\textwidth}
         \centering
         \includegraphics[width=\textwidth]{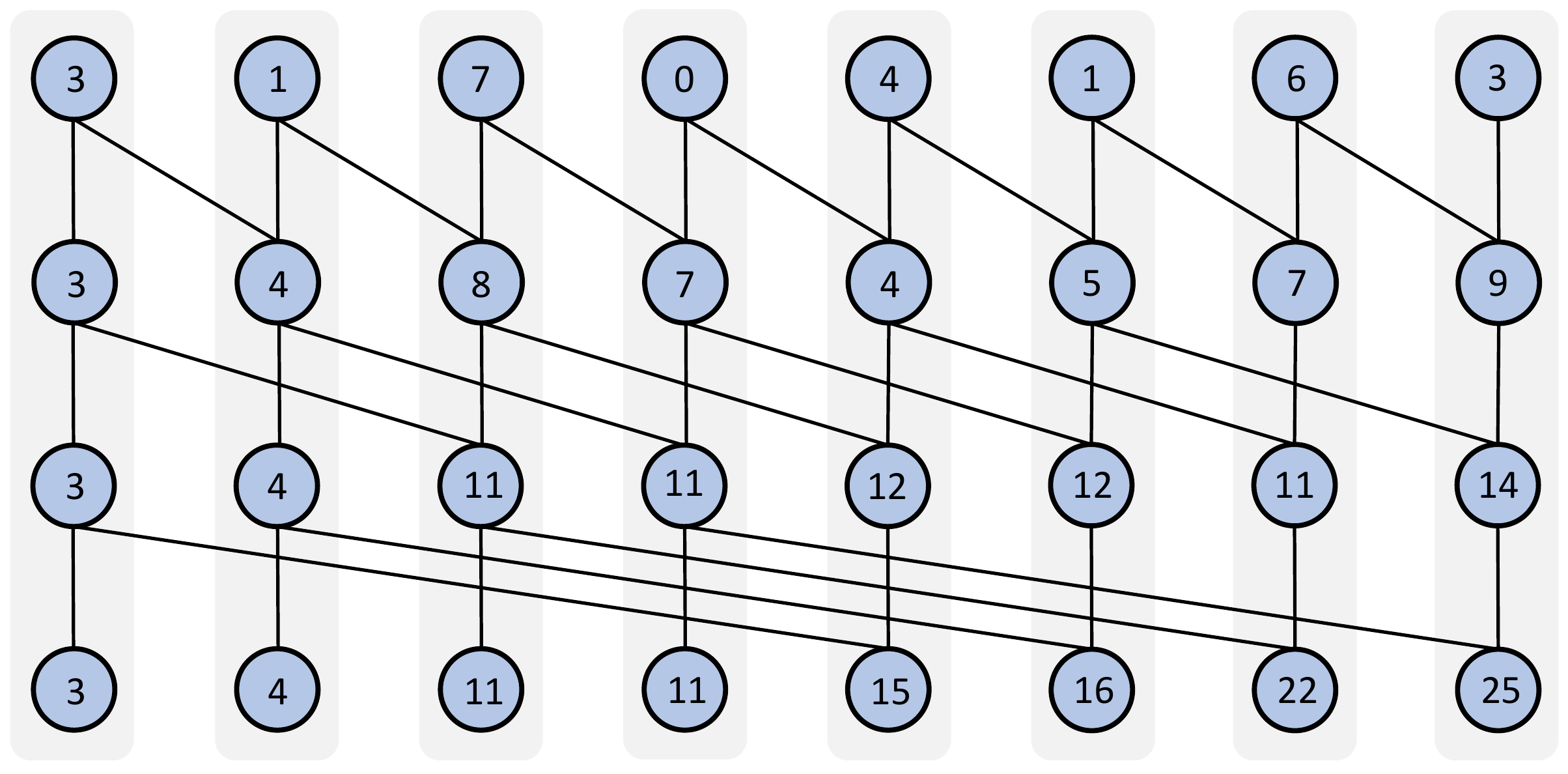}
         \caption{Naive scan}
         \label{fig:method_naive_scan}
     \end{subfigure}
     \hfill
     \begin{subfigure}[b]{0.47\textwidth}
         \centering
         \includegraphics[width=\textwidth]{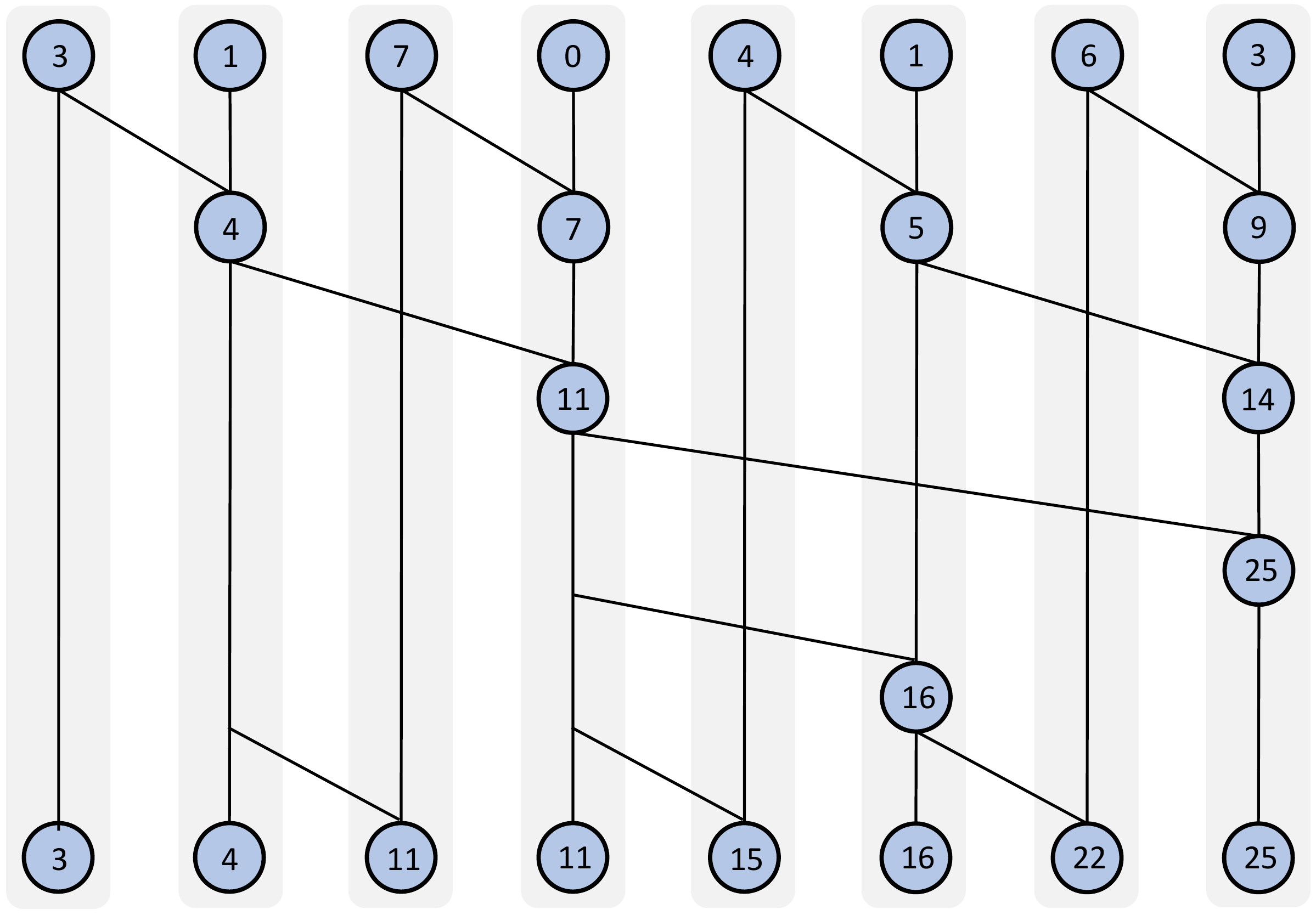}
         \caption{\textit{Reduce-then-scan}}
         \label{fig:method_reduce_then_scan}
     \end{subfigure}
        \caption{Tree-based parallel implementation of reduction and scan. Each grey box (column) represents a thread. Each thread runs a series of steps, and some steps must wait for results from other threads. Subfigures (a) and (b) show the naive binary tree-based approach for reduction and scan, respectively. Subfigure (c) presents a work-efficient scan algorithm using \textit{reduce-then-scan} strategy, which includes an up-sweep phase for accumulating the partial sums and a down-sweep phase for aggregating the prefix sums. The tree-based approach for reduction and scan generally requires much data communication between threads but this remains low latency in shared memory, and thus is suitable for GPU parallelization.}
        \label{fig:method_naive_parallel}
\end{figure}

The parallel versions of reduction and scan use a tree-based approach shown in Figure \ref{fig:method_naive_parallel}.
Note that effective parallelization of these types of binary tree traversals requires low latency sharing of partial sums across threads with appropriate synchronization, both aspects in which the large number of threads concurrently executing on the GPU greatly outperform multi-core CPU threads.

To obtain a parallel, work-efficient, and communication-avoiding prefix scan, we invoke the CUB library \citep{merrill2016single}.
The efficiency of their prefix-scan approaches a simple copy operation, as their prefix-scan requires the optimal \textasciitilde$2n$ data movements: $n$ reads and $n$ writes to the global memory.
The scan is constructed on two levels of organization: (1) a global device-wide scan and (2) a set of local block-wide scans within each thread block.
The local block-wide scan utilize a \textit{reduce-then-scan} strategy that can be visually resembled as an ``hourglass'' shape comprising an \textit{up-sweep} and a \textit{down-sweep} as shown in Figure \ref{fig:method_reduce_then_scan}.
In the up-sweep phase, we traverse the tree from leaves to root for computing the partial sums.
Then the running prefixes are aggregated in the block-wide down-sweep traversing back up the tree from root using the partial sums computed in the up-sweep phase.
The global scan implementation within CUB applies a single-pass chained-scan approach to achieve just \textasciitilde$2n$ global data movements.
The global scan further propose a decoupled look-back strategy by assigning each thread block a status flag indicating one of the three status: (1) aggregate (partial sum) of the block is available; (2) prefix of the block is available; and (3) no information about the block is available for other processor.
Then each block will perform the computation conditional on its predecessor's status flag.
We refer readers to \cite{merrill2016single} for more details on this algorithm.
In this paper, we extend these parallel algorithms for Cox models and Fine-Gray models based on the implementation of reduction and scan from the CUB library.

\subsection{GPU massive parallelization for parameter estimation}\label{2.7}

CCD is an inherently serial algorithm in which each iteration is based on the result of the last iteration.
As we mentioned earlier, even within an iteration $t$, CCD cycles through each covariate $j$ for $j = 1, 2, \ldots, P$ one by one and the computation work for the next covariate cannot begin until the current update finishes.
This serial algorithm however can still benefit greatly from parallelization by exploiting fine-grain problem decomposition within each iteration.
Within each iterate's covariate update, careful benchmarks reveal that over $95\%$ of the runtime lies in computing the log-likelihood gradient $g'(\beta_j)$ and Hessian $g''(\beta_j)$.

To understand the computational work, let $\bm{\delta} = (\delta_1, \ldots, \delta_N)\transpose$ be an N-dimensional column vector and $\mathbf{M}$ be an $N \times N$ loading matrix with entries
\begin{align}
  M_{ij} =
    \begin{cases}
      1 & \text{for $j \in R_1(Y_i)$}\\
      0 & \text{otherwise}.
    \end{cases}
\end{align}
Recall that the risk set $R_1(Y_i)$ contains all observations that have an observed event time equalling or after $Y_i$.
Thus if we arrange the observations by their observed time $Y_i$ in decreasing order, matrix $\mathbf{M}$ is clearly a lower triangular, binary matrix, and matrix-vector multiplication $\mathbf{M}[\exp \left(\mathbf{X}\bbeta \right)]$ becomes a prefix sum over the elements in $\exp \left(\mathbf{X}\bbeta \right)$, where we define exponentiation (exp) as element-wise operation.
For example, given $Y_{i} > Y_{i'}$, the set $R(Y_i)$ consists of all the observations from $R(Y_{i'})$ and the set $\{t: Y_t \in [Y_{i'}, Y_i]\}$, then
$\sum_{r \in R(Y_i)}\exp \left(\mathbf{x}_r\transpose\bbeta\right) = \sum_{r \in R(Y_{i'})}\exp \left(\mathbf{x}_r\transpose\bbeta\right) + \sum_{r \in \{t: Y_t \in [Y_{i'}, Y_i]\}}\exp \left(\mathbf{x}_r\transpose\bbeta\right)$.
Making these substitutions in Equation \ref{cox_l}, we arrive at
\begin{align}
L_{\text{\tiny pseudo}}(\bbeta) =
\bm{\delta}\transpose\mathbf{X}\bbeta -
\bm{\delta}\transpose \log \{S_{\text{\tiny pre}}[ \exp \left(\mathbf{X}\bbeta \right)]\} ,
\end{align}
where we define forming the logarithm (log) as element-wise operation, and $S_{\text{\tiny pre}}[\bnu]$ as the prefix sum of arbitrary vector $\bnu$.
Then the unidimensional gradient and Hessians under the Cox proportional hazards model falls out as
\begin{align}
g'(\beta_j) &= 
\bm{\delta}\transpose\mathbf{X}_j -
\bm{\delta}\transpose
\mathbf{G}
 \text{ and} \label{cox_g}\\
g''(\beta_j) &= 
- \bm{\delta}\transpose
\left(\mathbf{H} - \mathbf{G} \mathbf{\times} \mathbf{G} \right)
,\label{cox_h}
\end{align}
where 
\begin{align}
\mathbf{G} &= \frac
{S_{\text{\tiny pre}}[\exp \left(\mathbf{X}\bbeta\right) \bm{\times} \mathbf{X}_j]}
{S_{\text{\tiny pre}}[\exp \left(\mathbf{X}\bbeta \right)]} \text{ and} \\
\mathbf{H} &= \frac
{S_{\text{\tiny pre}}[\exp \left(\mathbf{X}\bbeta\right) \bm{\times} \mathbf{X}_j \bm{\times} \mathbf{X}_j]}
{S_{\text{\tiny pre}}[\exp \left(\mathbf{X}\bbeta \right)]}
\end{align}
and vector $\mathbf{X}_j$ is the $j$-th column of $\mathbf{X}$. Note that we define here multiplication ($\times$) and division ($/$) as element-wise operations as well.

While matrix-vector multiplication involving $\mathbf{M}$ takes $\mathcal{O}(N^2)$ operations, identifying the cumulative structure reduces the time complexity to linear by calculating prefix sum $S_{\text{\tiny pre}}[\bnu]$ in series, and parallelization further reduces the time complexity to $\mathcal{O}(\log_2 N)$ due to parallel scan's tree-based structure \citep{harris2007parallel}.
Finally, the vector-vector multiplication involving $\mathbf{\delta}\transpose$ can be calculated as an element-wise multiplication in parallel in constant time and a reduction through a binary-tree in $\mathcal{O}(\log_2 N)$.

Under the Fine-Gray model, let $\mathbf{W} = (w_1, \ldots, w_N)\transpose$ be an N-dimensional column vector of precomputed censoring weights described in previous section, and $\mathbf{N}$ as an $N \times N$ loading matrix with entries
\begin{align}
  N_{ij} =
    \begin{cases}
      1 & \text{for $j \in R_2(Y_i)$}\\
      0 & \text{otherwise} .
    \end{cases}
\end{align}
Recall that $R_2(Y_i) = \{r : Y_r < Y_i \cap \epsilon_r = 2\}$ and $R_1(Y_i)$ and $R_2(Y_i)$ are disjointed, so $\mathbf{N}$ is an upper triangular, binary matrix and $\mathbf{N}[\exp \left(\mathbf{X}\bbeta \right) \bm{\times} \mathbf{W}]$ is a suffix sum if we arrange the observations by their observed time $Y_i$ in decreasing order.
Then making the following substitutions in Equation \ref{cox_g} and \ref{cox_h} yields the unidimensional gradient and Hessians under the Fine-Gray model
\begin{align}
\mathbf{G} &= \frac
{S_{\text{\tiny pre}}[\exp \left(\mathbf{X}\bbeta\right) \bm{\times} \mathbf{X}_j] + S_{\text{\tiny suf}}[\exp \left(\mathbf{X}\bbeta\right) \bm{\times} \mathbf{X}_j \bm{\times} \mathbf{W}]}
{S_{\text{\tiny pre}}[\exp \left(\mathbf{X}\bbeta \right)] + S_{\text{\tiny suf}}[\exp \left(\mathbf{X}\bbeta \right) \bm{\times} \mathbf{W}]} \text{ and} \label{fg_g}\\
\mathbf{H} &= \frac
{S_{\text{\tiny pre}}[\exp \left(\mathbf{X}\bbeta\right) \bm{\times} \mathbf{X}_j \bm{\times} \mathbf{X}_j] + S_{\text{\tiny suf}}[\exp \left(\mathbf{X}\bbeta \right) \bm{\times} \mathbf{X}_j \bm{\times} \mathbf{X}_j \bm{\times} \mathbf{W}]}
{S_{\text{\tiny pre}}[\exp \left(\mathbf{X}\bbeta \right)] + S_{\text{\tiny suf}}[\exp \left(\mathbf{X}\bbeta \right) \bm{\times} \mathbf{W}]}, \label{fg_h}
\end{align}
where we define $S_{\text{\tiny suf}}[\bnu]$ as suffix sum of vector $\bnu$.

It is worth noting that the risk set under the competing risk setting consists of two disjoint sets due to multiple types of event, thus a single pass scan furnishes neither the numerator nor denominator.
Instead, the numerator and denominator of Equation \ref{fg_g} and \ref{fg_h} can be computed through a forward (prefix) scan $S_{\text{\tiny pre}}[\bnu]$ plus a backward (suffix) scan $S_{\text{\tiny suf}}[\bnu]$ together \citep{kawaguchi2020scalable}.

In summary, we can decompose the calculation of the gradient and Hessian for $\beta_j$ in the following four sequential steps:
\begin{enumerate}
    \item Read in non-zero $x_{ij}$ for $i = 1, 2, \ldots, N$ and update $[\mathbf{X}\pmb{\beta}]_i$ as
		$$ [\mathbf{X}\pmb{\beta}]_i^{\text{(new)}} = [\mathbf{X}\pmb{\beta}]_i + x_{ij}\Delta \beta_j.$$
		Then perform three element-wise embarrassingly parallel transformations that read from the new estimate of $[\mathbf{X}\pmb{\beta}]$, and output
		$[\exp \left(\mathbf{X}\bbeta \right)]$,
		$[\exp \left(\mathbf{X}\bbeta\right) \bm{\times} \mathbf{X}_j]$ and
		$[\exp \left(\mathbf{X}\bbeta\right) \bm{\times} \mathbf{X}_j \bm{\times} \mathbf{X}_j]$.
		Here we should keep in mind that $\mathbf{X}$ is generally sparse such that many elements $x_{ij}$ are zeros, and exponentiation is an expensive operation in floating-point.
    \item Scans:
    \begin{enumerate}
        \item Under the Cox model, we perform three forward scans that take the three output vectors of (1) as input, and return
        $S_{\text{\tiny pre}}[\exp \left(\mathbf{X}\bbeta \right)]$, 
        $S_{\text{\tiny pre}}[\exp \left(\mathbf{X}\bbeta\right) \bm{\times} \mathbf{X}_j]$ and 
        $S_{\text{\tiny pre}}[\exp \left(\mathbf{X}\bbeta\right) \bm{\times} \mathbf{X}_j \bm{\times} \mathbf{X}_j]$.
        \item Under the Fine-Gray model, we perform three forward scans and three backward scans that take the three output vectors of (1) as input, and return 
        $S_{\text{\tiny pre}}[\exp \left(\mathbf{X}\bbeta \right)]$, 
        $S_{\text{\tiny pre}}[\exp \left(\mathbf{X}\bbeta\right) \bm{\times} \mathbf{X}_j]$, 
        $S_{\text{\tiny pre}}[\exp \left(\mathbf{X}\bbeta\right) \bm{\times} \mathbf{X}_j \bm{\times} \mathbf{X}_j]$,
        $S_{\text{\tiny suf}}[\exp \left(\mathbf{X}\bbeta \right) \bm{\times} \mathbf{W}]$,
        $S_{\text{\tiny suf}}[\exp \left(\mathbf{X}\bbeta\right) \bm{\times} \mathbf{X}_j \bm{\times} \mathbf{W}]$ and
        $S_{\text{\tiny suf}}[\exp \left(\mathbf{X}\bbeta\right) \bm{\times} \mathbf{X}_j \bm{\times} \mathbf{X}_j \bm{\times} \mathbf{W}]$.
    \end{enumerate}
    \item An element-wise transformation that takes the output vectors of $(2)$ as well as the indicator vector $\bm{\delta}$ as input, and outputs two new vectors 
    $\bm{\delta} \bm{\times} \mathbf{G}$ and
    $\bm{\delta} \bm{\times} \left(\mathbf{H} - \mathbf{G} \mathbf{\times} \mathbf{G} \right)$.
    \item Two reductions that take the output vectors of $(3)$ as input, and output two double summations 
    $\bm{\delta}\transpose \mathbf{G}$ and
    $g''(\beta_j) = \bm{\delta}\transpose
    \left(\mathbf{H} - \mathbf{G} \mathbf{\times} \mathbf{G} \right)$.
    Note that the first term $\bm{\delta}\transpose\mathbf{X}_j$ in gradient $g'(\beta_j)$ can be precomputed and the value does not change during CCD.
\end{enumerate}

Although parallel computing of the above numerical operations is much faster than serial evaluation, one crucial limitation is that a memory transaction involving reading or writing from global memory may take up to two orders of magnitude more time than a regular numerical operation applied to the value sitting in the limited number of on-chip registers within a GPU (or CPU for that matter) \citep{holbrook2020massive}.
In order to minimize memory transactions, we fuse several of our operations together for both our serial and parallel implementations.
First, we fuse the multiple scans in step (2) together as a tuple-scan, which takes a tuple of three vectors as input, and outputs a tuple of three (or six) vectors.
Note that in Fine-Gray model, we perform the forward tuple-scan and the backward tuple-scan on the same tuple of three vectors, but read the input in two opposite directions simultaneously.
Similarly, we can fuse the multiple reductions in step (4) as a tuple-reduction.
Since the element-wise transformations in step (3) take $\mathcal{O}(1)$ time with GPU parallelization, step (3) and step (4) can be regarded as a transformation-reduction.
Finally, since both scan and reduction parallelization share the same binary-tree structure, we further fuse steps (2) - (4) together into a single kernel.
Through fusion, for example, the output tuple from the scans never need to be written to global memory, nor read back for the later transformation-reductions.
The fused kernel saves $2/3$ of the reads and writes than executing three separated kernels.
It is also worth noting that the computational work of the gradient  and Hessian evaluation share a similar structure and even some component such as $\mathbf{G}$, such that we have fused the computation of gradient and Hessian together to circumvent unnecessary kernel overhead and facilitate data reuse of these intermediate terms.

To exploit the sparsity of the design matrix $\mathbf{X}$ , we parallelize the transformation in step (1) using a sparse CUDA kernel, which only reads in and processes the non-zero entries while keeping other entries as zeros all the time during CCD updates.
This sparse kernel saves data movement as well as reduces memory bandwidth requirements  significantly when $\mathbf{X}$ is sparse, which is common in real-world scenarios.

\begin{figure}[ht]
    \centering
    \makebox[\textwidth][c]{\includegraphics[width=1\textwidth]{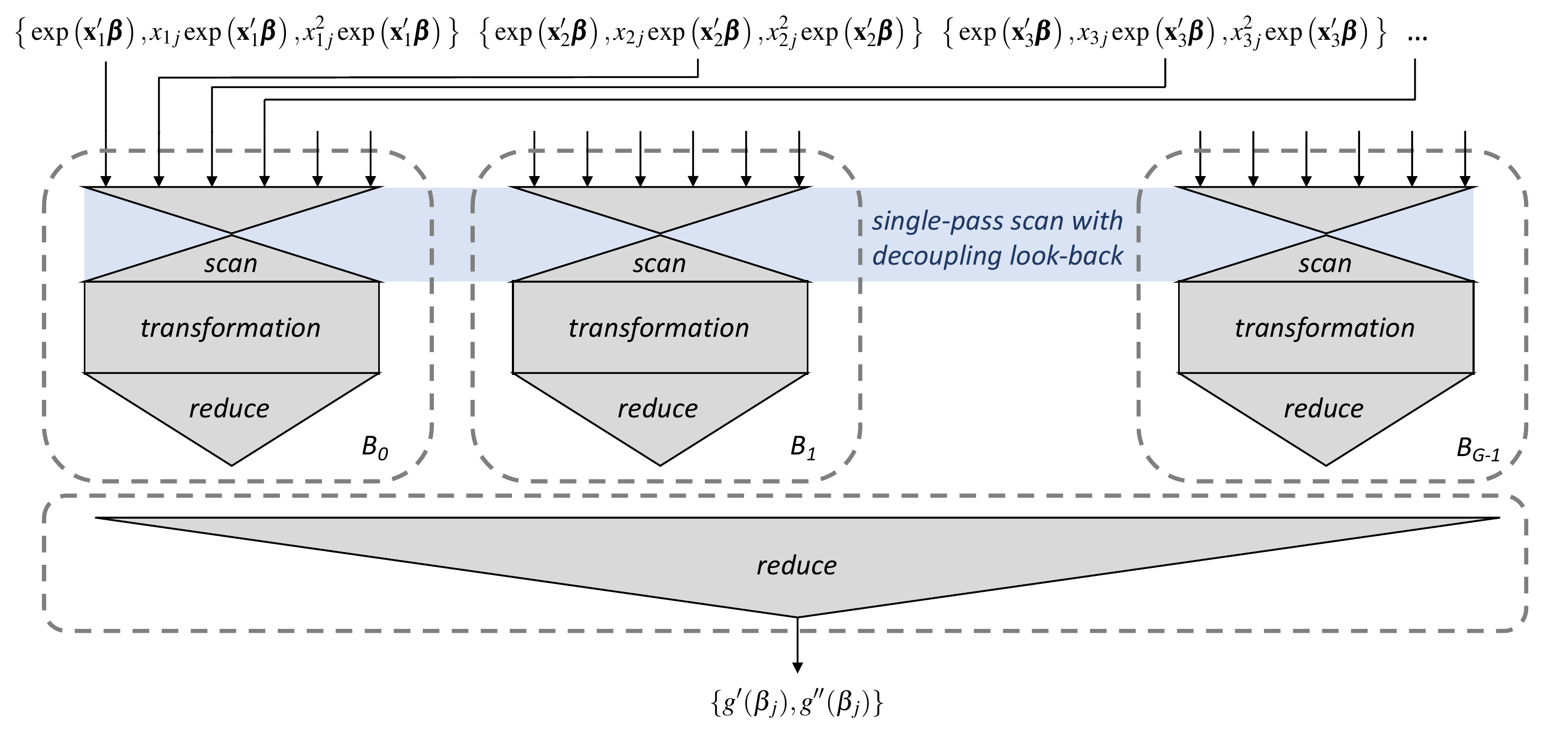}}
    \caption{Fused kernel for evaluating the gradient and Hessian for $\beta_j$. We fused the single-pass scan with decoupling look-back (represented by the ``hourglass''), the element-wise transformation (represented by the rectangle), and the reduction (represented by the upside down triangle) together in a fused kernel. Specifically, each of $G$ blocks (showed as a box in dashed line) reads a batch of triplets from global memory, performs a single-pass adaptive tuple-scan and a transformation-reduction to compute local partial sum of gradients and Hessians in shared memory, then efficiently adds the pair of partial sums in a binary-tree tuple-reduction within a single block and writes the resulting pair of gradient and Hessian to global memory.}
    \label{fig:method_fused}
\end{figure}

Figure \ref{fig:method_fused} illustrates our fused kernel for evaluating the log-likelihood gradient and Hessian on the GPU.
In this kernel, we spawn $S =  B \times IPT \times G$ threads, where $B$ is the number of concurrent threads forming a thread block, $IPT$ is the number of work-items (elements of input) evaluated per thread, and $G = \lceil \frac{N}{B \times IPT} \rceil$ denotes the number of thread blocks.
Both of the block size $B$ and the thread grain size $IPT$ are constrained by the hardware and are tunable constants.
In practice, we follow the parameter settings in CUB with $B = 128$ and $IPT = 15$ for the fused kernel.
For the sparse kernel we wrote, we choose $B = 256$ and $IPT = 1$.
In the figure, each box in dashed line represent a thread block and recall that
all threads within a block can access a shared memory that is a low-latency and on-chip \citep{nickolls2008scalable} and is useful for performing the efficient scan in step (2) and reduction in step (4).
The threads in parallel first read the values of tuple $\{\exp \left(\mathbf{x}_i\transpose\bbeta\right), x_{ij}\exp \left(\mathbf{x}_i\transpose\bbeta\right), x_{ij}^2\exp \left(\mathbf{x}_i\transpose\bbeta\right)\}$ for the current covariate column $j$ from global memory and then conduct the single-pass adaptive look-back tuple-scan \citep{merrill2016single} utilizing a \textit{reduce-then-scan} strategy.
Next, the threads read in the values of $\delta$ and perform the transformation with the on-chip cumulative sums and $\delta$.
The threads then perform a binary-tree tuple-reduction using shared memory, and one thread from each block writes its partial aggregates to global memory.
Finally, a single-block reduction kernel sums over the $G$ partial aggregates and writes the gradient $g'(\beta_j)$ and Hessian $g''(\beta_j)$ back to global memory.

\begin{figure}[h]
    \centering
    \makebox[\textwidth][c]{\includegraphics[width=1\textwidth]{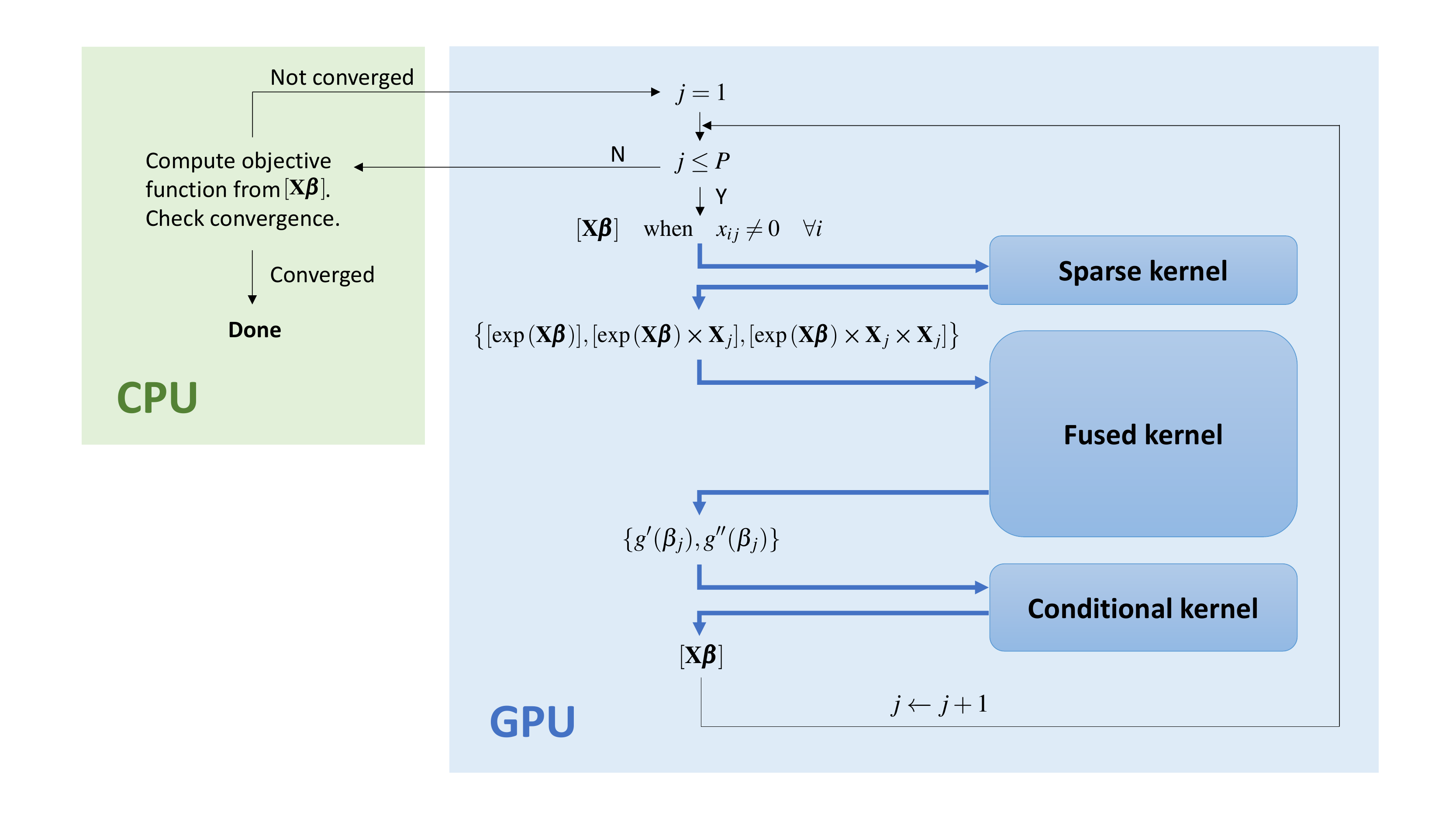}}
    \caption{The workflow of implementing CCD using GPU parallelization: for each $j = 1, \ldots, P$, a sparse kernel reads in the entries of $[\mathbf{X}\bbeta]$ for which $x_{ij} \neq 0$ and writes the updated tuple of vectors to the global memory of the GPU, then a fused kernel performs a tuple-scan and a transformation-reduction to compute the gradient and Hessian of the log-likelihood with respect to $\beta_j$.  Finally a conditional kernel computes $\Delta \beta_j$ and updates $\beta_j$ and $\mathbf{X}\bbeta$ if $\Delta \beta_j \neq 0$. Blue arrows represent data transactions to or from global memory.  No data transfer between the GPU and CPU is needed until CCD finishes a complete cycle.}
    \label{fig:method_ccd}
\end{figure}

When CCD processes the $j$-th covariate, we only need to update $\beta_j$ and corresponding vector $\mathbf{X}\bbeta$ if $\Delta \beta_j \neq 0$.
Recall that the computation of $\Delta \beta_j$ requires $g'(\beta_j)$ and $g''(\beta_j)$ when regularization applies.
Since both gradient $g'(\beta_j)$ and Hessian $g''(\beta_j)$ are computed on the GPU, we avoid $P$ data transfers between GPU and CPU in one CCD cycle by using a GPU kernel to  check if $\Delta \beta_j \neq 0$ and then update $\beta_j$ and $\mathbf{X}\bbeta$ if needed.
Figure \ref{fig:method_ccd} details the workflow to implement CCD using GPU parallelization.

It is worth noting that thread-divergent branches in a CUDA kernel can substantially impact performance as execution gets temporally serialized.
However, this is not an issue in our implementation because such branch divergence penalties only occur when threads within the same warp, but not across warps, need to execute alternative instructions.
The branches in scan and reduction kernel in CUB are mainly due to slightly different instructions for the first processing tile and the last processing tile.
Thus the divergence occurs across warps and does not have an impact on performance penalty.

\subsection{Multi-stream cross-validation} \label{2.8}

We use $k$-fold cross-validation to search for the optimum tuning parameters $\gamma$ or $\tau$ that controls the strength of regularization.
We search for different values for the tuning parameter.
For each value, the procedures are as follows:
\begin{enumerate}
	\item Randomly split data into $k$ partitions.
	\item For each of the $k$ partitions, we fit survival models via CCD on the remaining $k-1$ partitions and compute the predictive log-likelihood of this partition using the estimated $\hat{\bbeta}$.
	\item Average out-of-sample likelihood across all $k$ folds.
	\item Repeat Step (1) - Step (3) $r$ times to reduce spurious effects from random data partitioning.
\end{enumerate}
Finally, we select the tuning parameter with the smallest average out-of-sample likelihood as the desired optimal value.

We further improve the efficiency of cross-validation using multi-stream GPU and multi-threaded CPU approaches.
Here, a stream denotes a sequence of operations (kernels) that execute in issue-order on a GPU \citep{rennich2011cuda}.
Instead of the fitting $k$ partitioned models serially in a single (default) stream, we fit the $k$ models across $s$ streams in parallel, where $1 < s \leq k$ and each GPU stream is scheduled by an independent CPU thread.
Likewise, for the multi-threaded CPU approach, $s$ CPU threads evaluate the $k$ models in parallel.

\subsection{Comparison with an alternative massive parallelization of Cox models} \label{2.9}

In \cite{ko2022high}, the authors presented sample PyTorch code for parallelizing proximal gradient descent on modest-sized $\ell_1$ regularized dense Cox regression.
Here we provide a qualitative comparison of our method with this alternative approach, focusing on the per-cycle cost of cyclic coordinate descent and proximal gradient descent.

A single iteration of proximal gradient descent on $\ell_1-$regularized Cox regression requires the following steps:

\begin{enumerate}
	\item A matrix-vector multiplication $\mathbf{X}\pmb{\beta}$.
	\item An element-wise transformation $\exp{(\cdot)}$ on the output vector of (1).
	\item A scan on the output vector of (2).
	\item Two matrix-vector multiplications of $\mathbf{P}\pmb{\delta}$ and $\mathbf{X^T}[\pmb{\delta} - \mathbf{P}\pmb{\delta}]$, where the matrix $\mathbf{P} = (\pi_{ij})$ is defined as $\pi_{ij} = I(y_i \geq y_j)\exp{(x_i^T \beta)} / \sum_{i: y_i \geq y_j}\exp{(x_i^T \beta)}$.
\end{enumerate}

Without considering the specific parallel library, the above steps contain $\mathcal{O}(NP) + \mathcal{O}(N) + \mathcal{O}(N) + \mathcal{O}(N^2)$ operations, respectively.
Meanwhile, one sweep of coordinate descent implemented in Cyclops requires no more than $\mathcal{O}(NP)$ operations (the worst case is that the data is dense), as each of the four steps outlined in Section \ref{2.7} only at most requires $\mathcal{O}(N)$ operations.
Additionally, step (4) in the proximal gradient descent approach described earlier can also be reduced to $\mathcal{O}(N)$ by utilizing the same trick discussed in Section \ref{2.7}.

It is important to note that the number of iterations required to achieve an equivalent termination criterion is highly dependent on the data and is beyond the scope of our manuscript.

We would also like to emphasize that one of the main feature of our implementation is that it supports and benefits from a sparse data format when available, as discussed in Section \ref{2.7}.

\section{Results} \label{3}

We examine the performance of GPU vs CPU computing for fitting our massive sample-size survival models.
To accomplish this, we conduct a series of synthetic experiments to investigate the relative compute-time of our parallelization across different sample sizes.
We then reproduce a real-world study using our GPU implementation under a Cox model and extend the study to the competing risks setting.
If not specified, we use a system equipped with a 10 core 3.3 GHz Intel(R) Xeon(R) W-2155 CPU and an NVIDIA Quadro GV100 with 5120 CUDA cores and 32GB RAM that can achieve up to 7.4 Tflops double-precision point performance.

\subsection{Kernel fusion} \label{3.1}

Kernel fusion is one of the main strategy we apply for achieving optimal performance.
Taking the Cox model as an example, we compare three different implementations for steps (2) - (4) in Section \ref{2.7}:

\begin{enumerate}[label=(\alph*)]
\item Separated kernels
	\begin{enumerate}[label=(\roman*)]
    \item Three separated scans using \texttt{cub::DeviceScan::InclusiveSum},
    \item One transformation operation using \texttt{thrust::transform}, and
    \item Two separated reductions using \texttt{cub::DeviceReduce::Sum}.
    \end{enumerate}
\item Partially-fused kernels
	\begin{enumerate}[label=(\roman*)]
	\item A tuple-scan using \texttt{cub::DeviceScan::InclusiveScan}, and
	\item A transformed tuple-reduction using \texttt{cub::TransformInputIterator} and \texttt{cub::Devi\-ceReduce::Reduce}.
	\end{enumerate}
\item Single fused kernel
\end{enumerate}

\begin{figure}[H]
	\centering
	\includegraphics[width=0.7\textwidth]{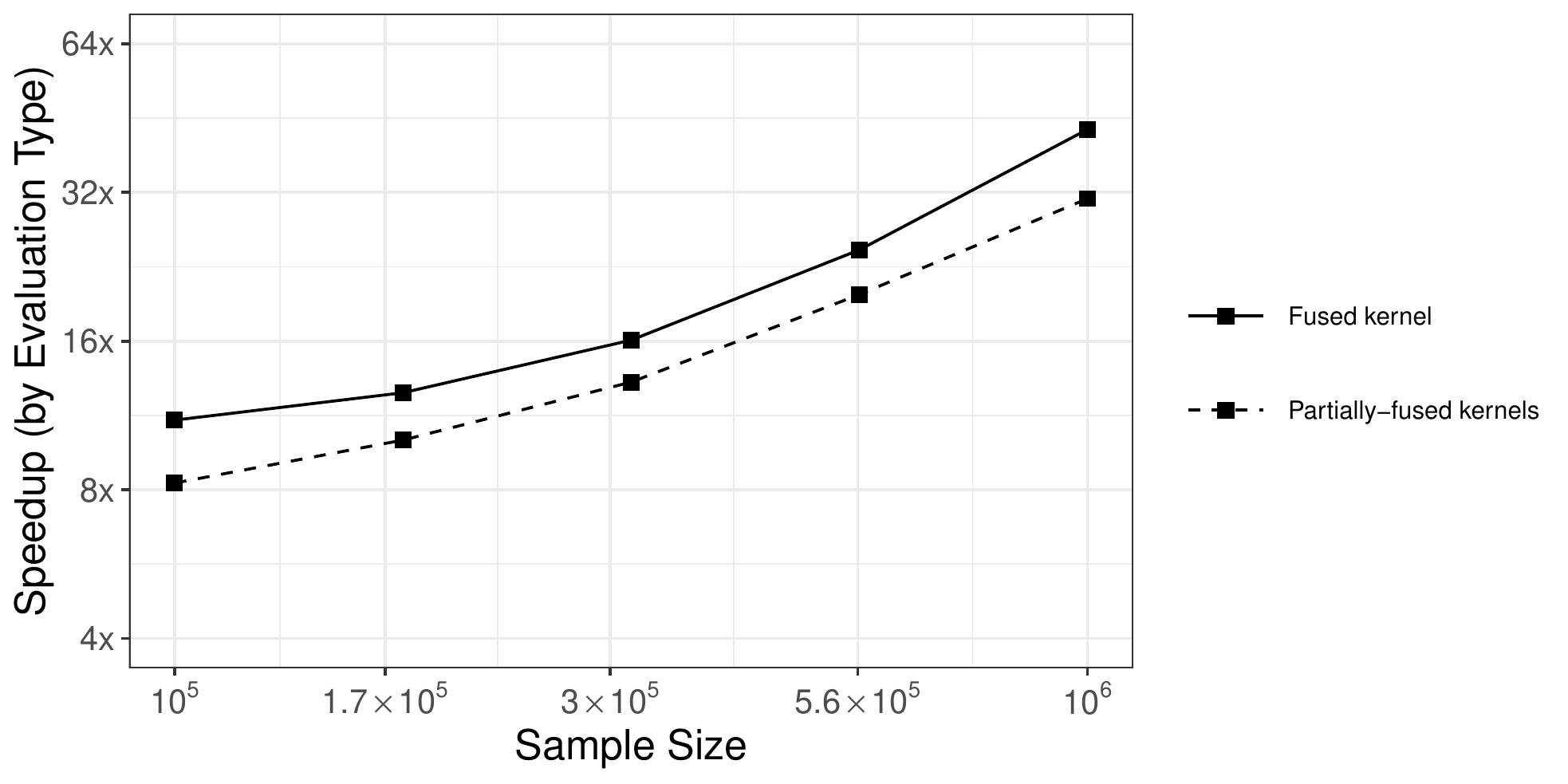}
	\caption{Speedup of the fused kernel and the partially-fused kernels over separated kernels for steps (2) - (4) in Section \ref{2.7}. The sample sizes range between $N = 10^5$ to $10^6$. Solid line shows the speedup of the fused kernel over separated kernels, and dashed line shows the speedup of the partially-fused kernel over separated kernels.}
	\label{fig:res_fused_speedup}
\end{figure}

We simulate the input vectors with $N = 100,000$ to $1,000,000$ from ${\rm Uniform}(1, 2)$.
Figure \ref{fig:res_fused_speedup} shows the speedup of the fused kernel and the partially-fused kernels over the separated kernels.
The partially-fused kernels are $8-31$ times faster than the separated kernels.
The single fused kernel further generates up to a 42-fold speedup compared with the separated kernels.
Table \ref{tb:res_fused} shows the runtimes (in milliseconds) in details when $N = 100,000$.
Interestingly, we find performing a tuple-scan on three vectors is almost three times faster than performing three separated scans ($0.0036$ ms v.s. $0.0098$ ms), demonstrating the values of fusion.

\begin{table}[ht]
\centering
\begin{tabular}{lrrr}
\hline
& \multicolumn{1}{c}{Separated} & \multicolumn{1}{c}{Partially-fused} & \multicolumn{1}{c}{Fused} \\
\multicolumn{1}{c}{Task} & \multicolumn{1}{c}{kernels} & \multicolumn{1}{c}{kernels} & \multicolumn{1}{c}{kernel} \\ \hline
Three scans    & 0.0098 & 0.0036                  & \multirow{3}{*}{\bigg\} 0.0052} \\
Transformation & 0.0415 & \multirow{2}{*}{\big\} 0.0035} &                         \\
Two reductions & 0.0066 &                         &                         \\ \hline
\multicolumn{1}{c}{\textbf{Total time}}
               & 0.0579 & 0.0071                  & 0.0052                  \\ \hline
\end{tabular}
\caption{Runtimes (in milliseconds) of the separated kernels, the partially fused kernels, and the fused kernel for steps (2) - (4) in Section \ref{2.7} when $N = 100,000$.}
\label{tb:res_fused}
\end{table}

\subsection{Synthetic experiments} \label{3.2}

We simulate indicator data $\mathbf{X}$ with $N = 100,000$ to $1,000,000$ samples and $P = 1000$ covariates with sparsity of $95\%$, where we randomly choose $5\%$ of the entries uniformly to be $1$s.
We then draw
\begin{align*}
    &\beta_j \sim N(0, 1) \times {\rm Bernoulli}(0.80)\ \forall\ j \text{ and} \\
    &T_i \sim {\rm Exponential}\left(\mathbf{x}_i \transpose \bbeta\right)\ \forall\ i,
\end{align*}
where the ${\rm Bernoulli}(0.80)$ specifies that, on average, $80\%$ of the $\beta_j$ are set to $0$ to induce sparsity, and $T_i$ represent the time-to-event time for individual $i$.
For generating competing risk times, we first draw:
\begin{align*}
    &\beta_{1j} \sim N(0, 1) \times {\rm Bernoulli}(0.80) \text{ and set }
    \beta_{2j} = - \beta_{1j}\ \forall\ j,
\end{align*}
where $\bbeta_1$ is the regression parameter of primary event and $\bbeta_2$ is the regression parameter of competing event \citep{kawaguchi2020scalable}.
Then we follow the design of Fine and Gray \citep{fine1999proportional}, where the cumulative incidence function (CIF) of primary event is a unit exponential mixture with mass $1-p$ at $\infty$ when $\mathbf{x}_i = \mathbf{0}$:
\begin{align*}
   \text{Pr}(T_{1i} \leq t_1 | \mathbf{x}_i) = 1 - \left[1 -p\{1-\exp(-t_1)\}\right]^{\exp(\mathbf{x}_i \transpose \bbeta_1)},
\end{align*}
and draw survival time of competing event $T_{2i}$ using an exponential distribution with rate $\exp(\mathbf{x}_i \transpose \bbeta_2)$.
We set $p = 0.5$ in practice.

\begin{figure}[H]
     \centering
     \begin{subfigure}[b]{0.8\textwidth}
         \centering
         \includegraphics[width=\textwidth]{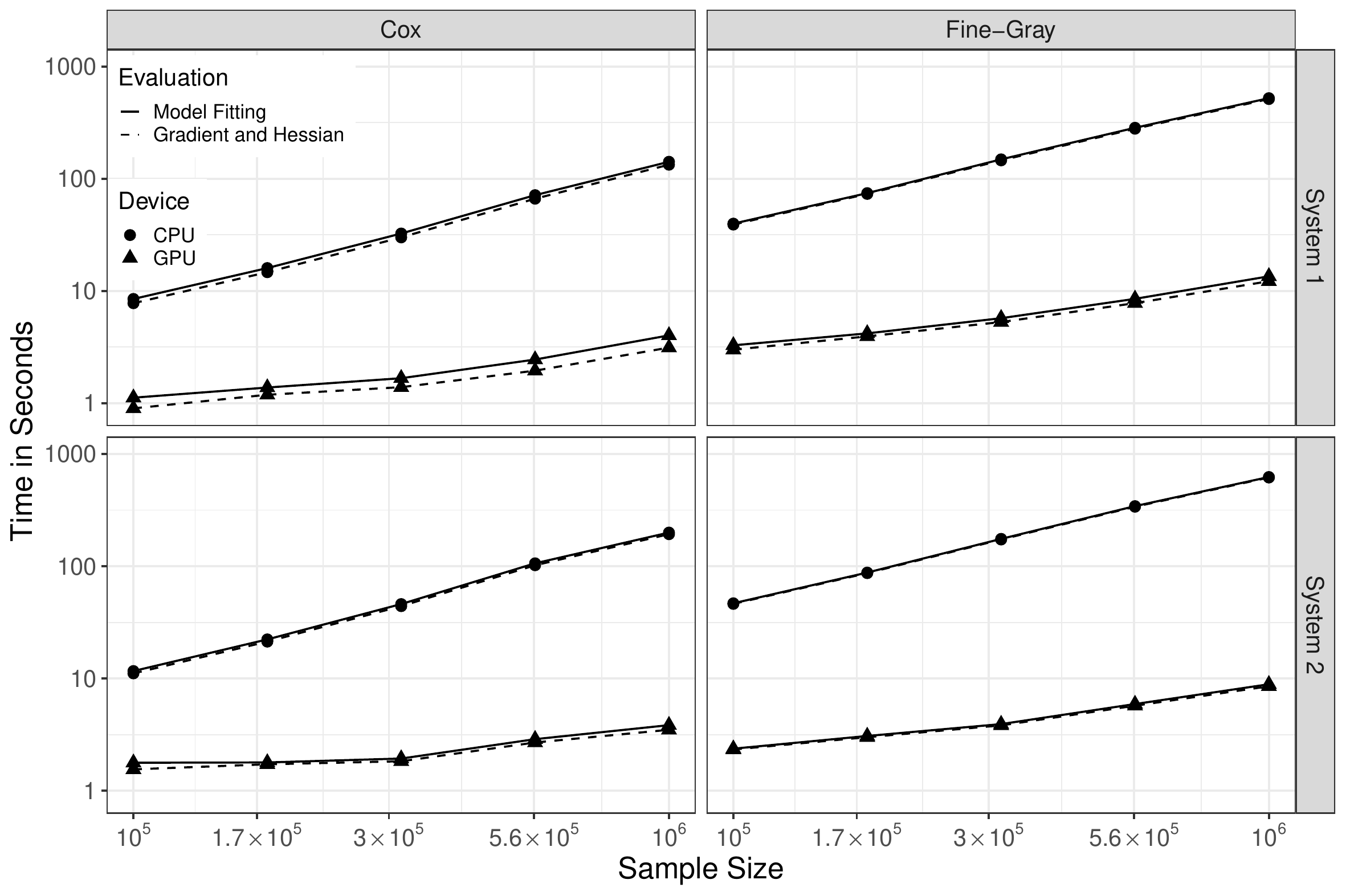}
         \label{fig:res_single_runtimes}
     \end{subfigure}
     \hfill
     \begin{subfigure}[b]{0.79\textwidth}
         \centering
         \includegraphics[width=\textwidth]{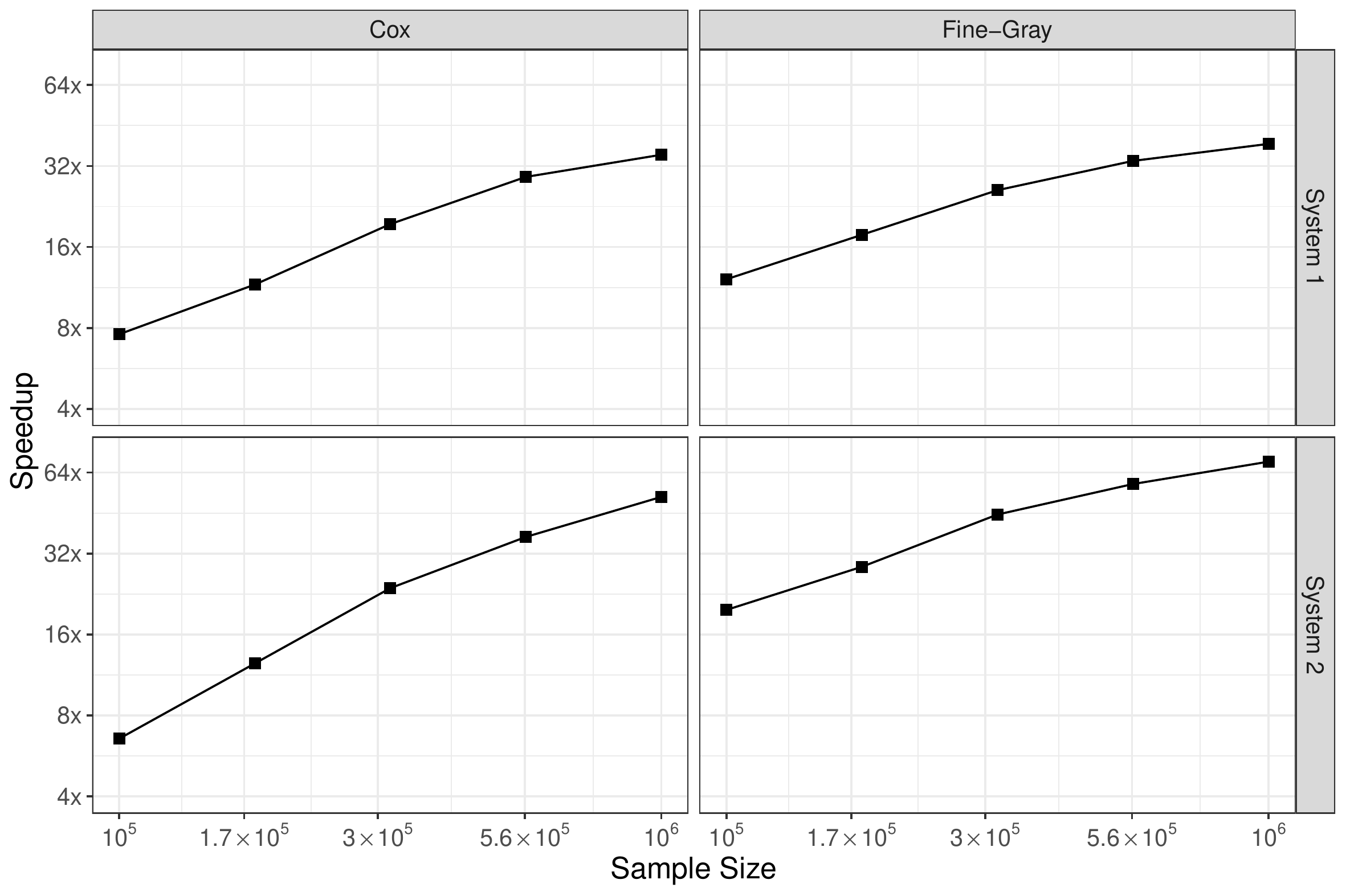}
         \label{fig:res_single_speedup}
     \end{subfigure}
        \caption{Runtimes (in seconds) and speedup of the GPU implementation relative to the CPU implementation for Cox and Fine-Gray models with a fixed $\ell_1$ penalty using a single GPU stream or CPU thread. The sample sizes range between $N = 10^5$ to $10^6$ with a sparsity of $95\%$. Solid lines show the total model fitting time, and dashed lines show the time for computing gradients and Hessians. The gap between the GPU's solid and dashed lines in the figures identify mostly computation that we did not port to the GPU, as data transfer between host and device during CCD updates accounts for less than $10\%$ of this overhead and $\sim1\%$ of the total model fitting time.}
        \label{fig:res_single}
\end{figure}

We fit these simulants under a fixed $\ell_1$ penalty with $\gamma = \sqrt{2}$ on a single CPU core and the default CUDA stream.
To provide a comprehensive comparison for the performance gains, we conduct the experiments on two systems with different technical specifications.
System 1 is equipped with a 3.3 GHz Intel(R) Xeon(R) W-2155 CPU (launch date 2017) and an NVIDIA Quadro GV100 (2018) with 5120 CUDA cores and 32GB RAM that can achieve up to 7.4 Tflops double-precision point performance.
System 2 is equipped with a 2.2 GHz Intel(R) Xeon(R) Silver 4214 CPU (2019) and an NVIDIA A100 (2020) with 6912 CUDA cores and 80GB RAM that can achieve up to 9.7 Tflops double-precision point performance.
Figure \ref{fig:res_single} presents runtimes comparisons across computing devices with a fixed number of covariates ($P = 1000$) with $95\%$ sparsity and varying sample size $N$.
We report both the total model fitting time and the time for computing gradients and Hessians, where the latter is our target of parallelization.
On system 1 which is equipped with a powerful CPU, we see that the GPU parallelization delivers up to a 42-fold speedup for both Cox and Fine-Gray models in terms of computing gradient and Hessians.
Despite additional data transfer and device initialization, GPU parallelization is still 35 $\times$ faster for this Cox model and 39 $\times$ faster for this Fine-Gray model relatively to our CPU implementation with one million samples.
On system 2 which is equipped with a more powerful GPU, we see that the GPU parallelization achieves up to a 52-fold speedup for this Cox model and a 70-fold speedup for this Fine-Gray model.
The data transfer between host and device during CCD updates only accounts for $\sim1\%$ of the total model fitting time.
We can see a rapid increase of runtimes on the CPU with increasing sample size, while the GPU approach continues to yield relatively shorter runtimes across varying sample sizes, as the devices is still not fully occupied.

\subsection{Multi-stream cross-validated experiments} \label{3.3}

\begin{figure}
	\centering
	\includegraphics[width=0.95\textwidth]{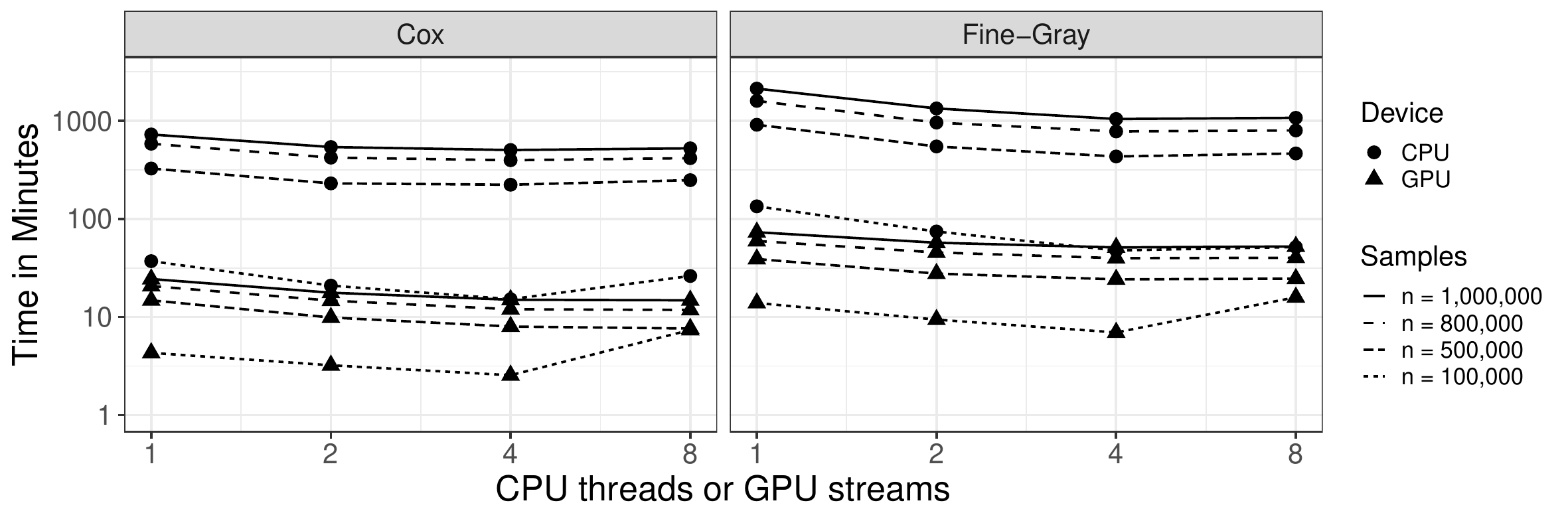}
    \caption{Runtimes for $\ell_1$ regularized Cox and Fine-Gray models with $10$-fold $10$-replicate cross-validation using multi-core CPU and multi-stream GPU computing. The sample sizes range between $N = 10^5$ to $10^6$ with a sparsity of $95\%$.}
    \label{fig:res_multi}
\end{figure}

We further use these synthetic experiments to explore the performance of our approach using multi-threaded CPU and multi-stream GPU computing by simultaneously searching for an optimal strength of regularization.
Here, we use $10$-fold cross-validation with $10$ repetitions per fold, resulting in $100$ cross-validation replicates to estimate an optimal $\gamma$ under $\ell_1$ regularization.
On the GPU, we distribute the $100$ replicates to $s$ CUDA streams driven by $s$ CPU threads.
We also allow each of the CPU threads to process the $100$ replicates directly on the CPU to demonstrate the performance of corresponding multi-threaded CPU parallelization.
Figure \ref{fig:res_multi} shows the runtimes of $\ell_1$ regularized Cox regression on varying number $s$ of threads or streams.
Generally, our parallelization achieves more speedup through multi-stream GPU computing than through multi-core CPU threads alone.
For example, the runtime of fitting a Cox model on $1,000,000$ samples reduces from nearly $9$ hours across $8$ CPU threads to $14$ minutes across $8$ CUDA streams.
We also observe that the curves flatten out as number of CUDA streams and CPU threads increases, and this pattern is particularly obvious in the experiments on smaller sample sizes and on the CPU.
This pattern suggests that there is both less computation to parallelize with relatively small sample sizes and that multi-core CPU parallelization remains limited by the smaller memory bandwidth available to the CPU.

\subsection{Cardiovascular effectiveness of antihypertensive drug classes} \label{3.4}

The large-scale evidence generation and evaluation across a network of databases for hypertension (LEGEND-HTN) study \citep{suchard2019comprehensive} provided real-world evidence on the comparative effectiveness and safety of five first-line antihypertensive drug classes using a retrospective, comparative new-user cohort design.
Specifically, LEGEND-HTN studied the relative risk of 55 health outcomes of interest , including three major cardiovascular events (acute myocardial infarction, hospitalization for heart failure, and stroke), six secondary effectiveness outcomes, and 46 safety outcomes.
Within each of nine observational health data sources, LEGEND-HTN employed propensity score matching or stratification for confounding adjustment and Cox proportional hazards modeling for hazard ratio (HR) estimation between new-users of each of the different drug classes.
Interestingly, LEGEND-HTN identified that new-users of thiazide or thiazide-like diuretics (THZs) have a lower risk as compared to new-users of angiotensin-converting enzyme inhibitors (ACEIs) in terms of several effectiveness and safety outcomes, even though ACEIs are the most commonly initiated monotherapy across databases.

\begin{table}
\centering
\begin{tabular}{p{4.5cm}rrLrrL}
\hline
  & \multicolumn{3}{l}{Before matching} & \multicolumn{3}{l}{After matching}\\
\cline{2-7}
 & THZ & ACEi & \multicolumn{1}{p{2cm}}{Standardised \newline difference} & THZ & ACEi & \multicolumn{1}{p{2cm}}{Standardised \newline difference} \\
\hline
\textbf{Age group (years)} & & \\
$10$-$14$                             & $0.1\%$ & $0.1\%$ & -0.02 & $0.1\%$ & $0.1\%$ & -0.01\\
$15$-$19$                             & $0.4\%$ & $0.6\%$ & -0.02 & $0.5\%$ & $0.5\%$ & 0\\
$20$-$24$                             & $1.5\%$ & $1.0\%$ & 0.04 & $1.4\%$ & $1.2\%$ & 0.02\\
$25$-$29$                             & $3.3\%$ & $2.1\%$ & 0.07 & $3.0\%$ & $2.8\%$ & 0.01\\
$30$-$34$                             & $5.5\%$ & $3.8\%$ & 0.08 & $5.1\%$ & $4.9\%$ & 0.01\\
$35$-$39$                             & $7.7\%$ & $6.0\%$ & 0.07 & $7.3\%$ & $7.2\%$ & 0\\
$40$-$44$                             & $10.1\%$ & $8.5\%$ & 0.05 & $9.8\%$ & $9.9\%$ & 0\\
$45$-$49$                             & $12.2\%$ & $11.4\%$ & 0.03 & $12.1\%$ & $12.1\%$ & 0\\
$50$-$54$                             & $13.8\%$ & $14.0\%$ & -0.01 & $13.9\%$ & $13.8\%$ & 0\\
$55$-$59$                             & $13.0\%$ & $14.5\%$ & -0.05 & $13.3\%$ & $13.5\%$ & -0.01\\
$60$-$64$                             & $10.6\%$ & $12.5\%$ & -0.06 & $11.0\%$ & $10.9\%$ & 0\\
$65$-$69$                             & $8.1\%$ & $9.6\%$ & -0.05 & $8.4\%$ & $8.5\%$ & 0\\
$70$-$74$  
   & $5.5\%$ & $6.6\%$ & -0.05 & $5.7\%$ & $5.9\%$ & -0.01\\
$75$-$79$                             & $4.5\%$ & $4.9\%$ & -0.02 & $4.6\%$ & $4.7\%$ & 0\\
$80$-$84$                             & $3.1\%$ & $3.5\%$ & -0.02 & $3.2\%$ & $3.4\%$ & -0.01\\
$85$-$89$                             & $0.7\%$ & $0.8\%$ & 0.02 & $0.7\%$ & $0.6\%$ & 0.01\\
\hline
\textbf{Gender} & & \\
Female                                & $63.3\%$ & $44.2\%$ & 0.39 & $60.9\%$ & $61.6\%$ & -0.02\\
\hline
\multicolumn{7}{l}{\textbf{Medical history (general)}} \\
Chronic obstructive \newline lung disease      & $3.1\%$  & $3.6\%$ & -0.03 & $3.1\%$ & $3.3\%$ & -0.02\\
Diabetes                              & $7.3\%$  & $24.2\%$ & -0.48 & $7.8\%$ & $8.2\%$ & -0.02\\
Hyperlipidemia                        & $30.8\%$ & $42.4\%$ & -0.24 & $32.3\%$ & $32.0\%$ & 0.01\\
Obesity                               & $16.3\%$ & $13.8\%$ & 0.07 & $15.4\%$ & $15.4\%$ & 0\\
Osteoarthritis                        & $11.5\%$ & $11.2\%$ & 0.01 & $11.5\%$ & $11.9\%$ & -0.01\\
\hline
\multicolumn{7}{l}{\textbf{Medical history (cardiovascular disease)}} \\
Cerebrovascular disease               & $1.5\%$  & $2.2\%$ & -0.06 & $1.5\%$ & $1.6\%$ & -0.01\\
Coronary arteriosclerosis             & $2.0\%$  & $3.8\%$ & -0.11 & $2.0\%$ & $2.1\%$ & 0\\
Heart disease                         & $8.0\%$  & $10.7\%$ & -0.09 & $8.0\%$ & $8.2\%$ & -0.01\\
\hline
\textbf{Medication use} & & \\
Anti-inflammatory and \newline antirheumatic products & $47.8\%$ & $45.6\%$ & 0.04 & $46.9\%$ & $47.6\%$ & -0.01\\
Antithrombotic agents                        & $17.0\%$ & $24.4\%$ & -0.18 & $17.4\%$ & $17.7\%$ & -0.01\\
Beta blocker                                 & $14.1\%$ & $20.5\%$ & -0.17 & $14.5\%$ & $14.5\%$ & 0\\
\hline
\end{tabular}
\caption{Baseline hypertensive patient characteristics for new-user of THZ and ACEi in the Optum EHR database. We conduct a propensity score matching with the matching ratio of $1:1$. Less standardised difference of population proportions after matching indicate improved balance between two new-user cohorts in terms of confounders. THZ = thiazide or thiazide-like diuretics. ACEi = angiotensin-converting enzyme inhibitors.}
\label{tb:res_patient}
\end{table}

\begin{figure}
	\centering
	\includegraphics[width=0.89\textwidth]{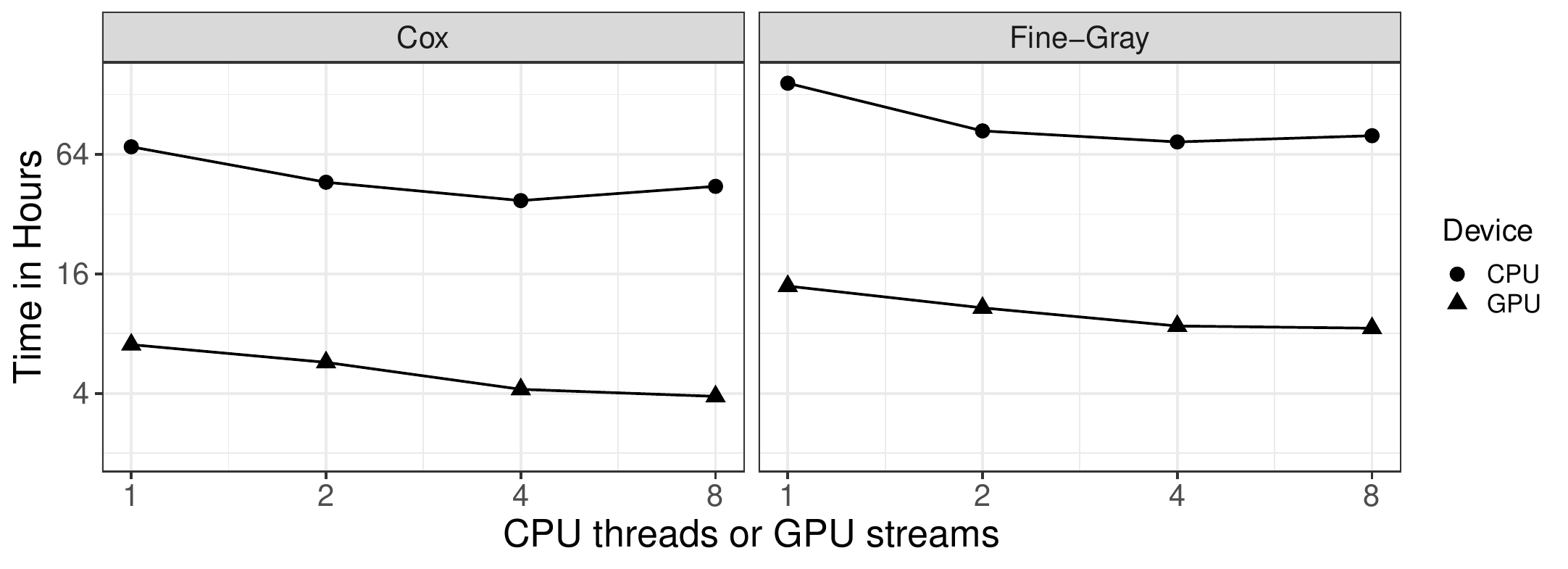}
    \caption{Runtimes (in hours) for $\ell_1$ regularized Cox and Fine-Gray models using multi-stream GPU and multi-core CPU computing. The data contain $434,866$ new-users of THZs and ACEIs, each with $9,811$ baseline patient characteristics covariates. We add a $\ell_1$ penalty on all baseline covariates and use a $10$-fold $10$-replicate cross-validation to search for optimum tuning parameters.}
	\label{fig:res_legend}
\end{figure}

Here we examine patients initiating ACEIs and THZs, where the outcome is hospitalization for heart failure from the Optum\textsuperscript{\tiny\textregistered} de-identified Electronic Health Record dataset (Optum EHR).
Optum EHR represents data from 85 million individuals that are commercially or Medicare insured in the United States. 
The data contain medical claims, pharmacy claims, laboratory tests, vital signs, body measurements, and information derived from clinical notes.
A total of $1,014,618$ patients are included in our study, $75.8\%$ of whom initiated an ACEI and $24.2\%$ of whom initiated a THZ.
We consider the main treatment covariate (ACEI or THZ exposure), in addition to $9,811$ baseline patient characteristic covariates.
Table \ref{tb:res_patient} presents a small selection of major patient baseline characteristics.
From all baseline characteristics, we build a propensity score model and match ACEI and THZ new-users in a $1:1$ ratio.
A total of $434,866$ patients were kept for further analysis after propensity score matching.
We find ACEI new-users are more likely to be male, have diabetes, hyperlipidemia or heart disease comparing with patients initiating a THZ before propensity score matching.
The THZ and ACEi cohorts are well-balanced on all $9,811$ baseline patient characteristics after matching, identified through low standardized difference of population proportions.
The average sparsity of patient characteristic covariates is $97.11\%$, which means $2.89\%$ of the entries are non-zero, occupying about $2.96$GB RAM in sparse matrix format.
We first fit a Cox proportional hazards model to estimate the HR between THZ and ACEi initiation for the risk of hospitalization for heart failure.
We also fit a Fine-Gray model in which we consider acute myocardial infarction as a competing risk of hospitalization for heart failure, given myocardial infarction substantially elevates the future risk of heart failure.
We include all patient characteristics as additional covariates in the Cox and Fine-Gray models with $\ell_1$ regularization on all variables except the treatment variable to achieve a limited form of adjustment for possible residual confounding, and again employ a 10-fold 10-replicate cross-validation to search for optimal tuning parameters.
The analyses require almost two days to fit the regularized Cox model and more than three days to fit the regularized Fine-Gray model using eight CPU cores, while only taking $3.87$ hours and $8.57$ hours for the regularized Cox model and the regularized Fine-Gray model with our GPU implementation.
Figure \ref{fig:res_legend} reports the runtimes for regularized Cox and Fine-Gray models on varying numbers of CUDA streams versus the corresponding multi-core CPU parallelization.

Through massive parallelization, we find that initializing with a THZ shows better effectiveness than initializing with an ACEI in terms of  hospitalization for heart failure risk under both the Cox model (HR $0.83$, $95\%$ bootstrapped percentile interval [BPI] $0.72-0.94$) and the Fine-Gray model (HR $0.83$, $95\%$ BPI $0.71-0.95$),
where we provide the BPIs as crude measures of sampling variability given the challenges in constructing nominal confidence intervals for regularized models \citep{casella2010penalized, chatterjee2011bootstrapping}.
There are 215 and 134 out of 9,811 baseline covariates with non-zero effect sizes in the Cox and Fine-Gray models, respectively; these covariates include age, gender, hyperlipidemia, diabetes, osteoarthritis, and heart disease.
While our estimates remain in line with the LEGEND-HTN study, they are reassuring in two ways.
First, massive parallelization has enabled us to provide additional adjustment for possible residual confounding due to unobserved imbalance after propensity score matching without overfitting through cross-validation.
Second, a consistent estimate after controlling for an obvious competing risk reduces concern over bias from informative censoring.
Neither including thousands of baseline covariates nor handling a competing risk at this scale were possible in the original LEGEND-HTN study due to their erroneous time requirements; both are now feasible.

\section{Discussion}

This paper presents a time- and memory-efficient GPU implementation of regularized Cox and Fine-Gray regression models for analyzing large-scale time-to-event data with competing risks.
We efficiently implement it in the open-source \texttt{R} package \texttt{Cyclops}  \citep{suchard2013massive}.
In simulation studies, our GPU implementation is $35-70$ times faster than the equivalent CPU implementation with up to 1 million samples.
In our real-world example with $\sim$ 400,000 hypertension patients and $\sim$ 9,000 covariates, massive parallelization reduces the total runtimes of both regularized Cox regression and regularized Fine-Gray regression with cross-validation from a few days on multi-core CPUs to few hours on a GPU.

The observed speed-up is a combination of algorithmic advances as well as hardware optimization. 
First, we observe that the cumulative structure of the risk set can be computed by a single-pass parallel scan algorithm, which can be very efficiently computed in parallel using a tree-based approach.
This enables us to develop a work efficient and communication-avoiding tuple-scan algorithm in accumulating statistics about individuals in the risk set, dramatically speeding up likelihood, gradient, and Hessian evaluations.
In particular, this tree-based scan algorithm takes great advantage of the high-speed shared memory of thread block on GPUs. 
Furthermore, as CCD is an inherently serial algorithm, we fuse multiple serial steps (1) transformations, (2) scans and (3) reductions together into a single vectorizable kernel.
This fusion is possible because transformation is relatively lightweight operation, and scan and reduction algorithm share the same binary-tree structure.
This kernel fusion reduces expensive memory transactions and kernel overheads which usually prevent the serial algorithm from benefiting by parallel computing.
We also exploit the sparsity of the design matrix $\mathbf{X}$ through a sparse CUDA kernel to further save data movements.
Finally, we only off-load the most computationally intensive routines to the GPU without rewriting the whole codebase.

There are numerous opportunities for improvement.
For instance, our multi-stream implementation for $k$-fold cross validation has not achieved full concurrency for $>1$ replicates, especially for relatively small sample size ($\sim$ 100,000) due to the low GPU utilization.
The other potential problem is that we have to replicate the design matrix on each CUDA stream, which increases the memory bandwidth requirement.
To overcome these limitations, one may use a single larger kernel to perform many repetitions of $k$-fold cross-validation on a single stream.
By increasing the computational work in a single kernel, we can over-subscribe GPU threads to achieve a higher utilization and improve data reuse, but of course this creates the danger of exhausting register memory.
In addition, we plan to add inference on parameter estimates via bootstrapping by synchronizing the computation of sub-samples.
Both bootstrapping and cross-validation require independent computations on different sub-samples, which may benefit from parallel computing.

Finally, GPU hardware and libraries are rapidly maturing, and statisticians stand to benefit from GPU parallelization.
We conclude this work with a few tips for educational purposes.
First, simple building blocks such as scans and reductions are ubiquitous in statistical inference.
One can easily apply cutting-edge parallel implementations of these building blocks to gain instant speed-up.
Second, kernel fusions should be employed whenever possible for serial algorithms to reduce expensive memory transactions and kernel overheads.
Finally, when optimize an existing program, one should consider off-loading the most computationally intensive routines to the GPU in stages, before rewriting the whole codebase.

\section{Acknowledgments}
This research was supported through US National Institutes of Health grant R01 AI153044 and we gratefully acknowledge support from NVIDIA Corporation with the donation of parallel computing resources.

\section{Disclosures}
MJS is an employee and share-holder of Johnson \& Johnson.  The remaining authors report there are no competing interests to declare.

\bigskip
\begin{center}
{\large\bf SUPPLEMENTARY MATERIALS}
\end{center}

\begin{description}

\item[Cyclops:] R-package Cyclops containing code to perform the analysis described in the article is available at \url{https://github.com/OHDSI/Cyclops/tree/gpu_cox}.

\item[R scripts for executing the experiments:]
R scripts for executing the experiments in Section~ 3. are available at \url{https://github.com/suchard-group/gpu_survival_analysis_supplement}. Unfortunately, we are unable to provide access to the Optum de-identified Electronic Health Records due to data licensing constraints. However, we provide the cohorts definition that can reproduce our real-world analysis for reader who do license the Optum de-identified Electronic Health Records data source.

\end{description}



\bibliographystyle{natbib}
\bibliography{reference.bib}
\end{document}